\documentclass[a4paper,preprintnumbers,11pt,tikz]{article}
\pdfoutput=1
\usepackage{jheppub}
\usepackage{amsmath,amssymb,bm,graphicx,bbold,epsf,colordvi}
\usepackage{lipsum}
\usepackage{makecell}
\usepackage{soul}
\usepackage{braket}
\usepackage{bm}
\usepackage{appendix}
\allowdisplaybreaks 
\usepackage{color}
\usepackage[dvipsnames]{xcolor}

\usepackage{mathrsfs}
\usepackage{slashed}
\allowdisplaybreaks
\usepackage[abs]{overpic}
\usepackage[export]{adjustbox}
\usepackage{bm}
\usepackage{scalerel}
\usepackage{accents}

\newcommand{\bef}{\begin{figure}[hbt]\centering}
\newcommand{\eef}{\end{figure}}
\usepackage{mathtools}
\usepackage{subfigure}
\usepackage{booktabs}
\usepackage{graphicx}
\graphicspath{ {./figure/} }
\usepackage{comment}
\usepackage{lipsum}

\newcommand{\beq}{\begin{equation}}
\newcommand{\eeq}{\end{equation}}
\def\bea#1\eea{\begin{align}#1\end{align}}

\newcommand{\R}{{\mathcal R}}

\def \be  {\begin{equation}}
\def \ee  {\end{equation}}
\def \ba  {\begin{eqnarray}}
\def \ea  {\end{eqnarray}}

\newcommand{\nn}{\nonumber}

\allowdisplaybreaks

\usepackage{tikz}
\usepackage{physics}
\usepackage{xcolor}
\usetikzlibrary{calc}
\tikzset{>=latex} 
\usetikzlibrary{decorations.pathreplacing} 

\colorlet{myblue}{blue!70!black}
\colorlet{mydarkblue}{blue!40!black}
\colorlet{mygreen}{green!40!black}
\colorlet{myred}{red!65!black}
\tikzstyle{vector}=[->,very thick,myblue,line cap=round]
\tikzstyle{ptmiss}=[->,dashed,thick,myred,line cap=round]
\tikzstyle{cone}=[thin,blue!50!black,fill opacity=0.8]

\newcommand\jetcone[4]{
  \pgfmathanglebetweenpoints{\pgfpointanchor{#1}{center}}{\pgfpointanchor{#2}{center}}
  \edef\tmpang{\pgfmathresult}
  \coordinate (tmpO) at ($(#1)+(\tmpang:0.02)$); 
  \coordinate (tmpC) at ($(#2)+(\tmpang-180:{abs(#4)+0.4})$); 
  \coordinate (tmpL) at ($(tmpC)+(\tmpang+90:#3)$); 
  \coordinate (tmpR) at ($(tmpC)+(\tmpang-90:#3)$); 
  \draw[thin,blue!50!black,fill opacity=0.8, 
        top color=blue!50!black!50,bottom color=blue!50!black!60,shading angle=\tmpang,rotate=\tmpang]
    (tmpC) ellipse({#4} and {#3});
  \begin{scope}
    \clip[rotate=\tmpang] (tmpR) -- (tmpO) -- (tmpL) arc(90:-90:{#4+0.6} and {#3});
    \draw[vector] (tmpO) -- (#2);
  \end{scope}
  \draw[thin,blue!50!black,fill opacity=0.8, 
        top color=blue!50!black!30,bottom color=blue!40!black!50,shading angle=\tmpang,rotate=\tmpang]
    (tmpL) arc(90:270:{#4} and {#3}) -- (tmpO) -- cycle;
}

\makeatletter
\def\@fpheader{~}
\makeatother

\usepackage[Q=yes,pverb-linebreak=no]{examplep}

\title{QCD resummation of dijet azimuthal decorrelations in pp and pA collisions}

\author[a]{Mei-Sen Gao}
\author[b,c,d]{, Zhong-Bo Kang}
\author[a,e,f]{, Ding Yu Shao}
\author[g]{, John Terry}
\author[a]{and Cheng Zhang}

\affiliation[a]{Department of Physics and Center for Field Theory and Particle Physics, Fudan University, Shanghai, China}
\affiliation[b]{Department of Physics and Astronomy, University of California, Los Angeles, CA 90095, USA}
\affiliation[c]{Mani L. Bhaumik Institute for Theoretical Physics, University of California, Los Angeles, CA 90095, USA}
\affiliation[d]{Center for Frontiers in Nuclear Science, Stony Brook University, Stony Brook, NY 11794, USA}
\affiliation[e]{Key Laboratory of Nuclear Physics and Ion-beam Application (MOE), Fudan University, Shanghai, China}
\affiliation[f]{Shanghai Research Center for Theoretical Nuclear Physics, NSFC and Fudan University, Shanghai 200438, China}
\affiliation[g]{Theoretical Division, Los Alamos National Laboratory, Los Alamos, NM 87545, USA}

\emailAdd{msgao@fudan.edu.cn,zkang@ucla.edu,dingyu.shao@cern.ch,jdterry@lanl.gov, chengzhang\_phy@fudan.edu.cn}

\abstract
{We study the azimuthal angular decorrelations of dijet production in both proton-proton (pp) and proton-nucleus (pA) collisions. By utilizing soft-collinear effective theory, we establish the factorization and resummation formalism at the next-to-leading logarithmic accuracy for the azimuthal angular decorrelations in the back-to-back limit in pp collisions. We propose an approach where the nuclear modifications to dijet production in pA collisions are accounted for in the nuclear modified transverse momentum dependent parton distribution functions (nTMDPDFs), which contain both collinear and transverse dynamics. This approach naturally generalizes the well-established formalism related to the nuclear modified collinear parton distribution functions (nPDFs). We demonstrate strong consistency between our methodology and the CMS measurements in both pp and pA collisions, and make predictions for dijet production in the forward rapidity region in pA collisions at LHC kinematics and for mid-rapidity kinematics at sPHENIX. Throughout this paper, we focus on the application of this formalism to a simultaneous fit to both collinear and transverse momentum dependent contributions to the transverse momentum dependent distributions. 
}

\begin{document}
\preprint{LA-UR-23-24761}
\maketitle

\section{Introduction}

\label{sec:intro}

The investigation of high-energy proton-proton (pp) and proton-nucleus (pA) collisions is a crucial area of study in particle and nuclear physics, as it provides valuable insight into the fundamental structure of matter and the strong interaction among its constituents~\cite{Accardi:2012qut,Albacete:2013ei,Aschenauer:2016our,Belmont:2023fau}. Jet production is a crucial observable in these collisions, where collimated sprays of particles produced by the strong force, described by quantum chromodynamics (QCD), are observed. One of the key features of jet production in proton-proton and proton-nucleus collisions is the azimuthal angular distribution, or the difference in the azimuthal angle between the two jets. In the perturbative region, this decorrelation is a result of emissions from both the initial and final states that can alter the direction of the jets. The study of azimuthal decorrelation is critical for a deeper understanding of QCD jets and for testing QCD predictions and searching for new physics.

When one studies the dijet pseudorapidity spectrum while integrating over the full range of the azimuthal angle, the observable can be studied within the usual collinear factorization~\cite{Collins:1989gx} and such a pseudorapidity spectrum is directly sensitive to the collinear parton distribution functions (PDFs), allowing us to constrain longitudinal motion of partons inside a free nucleon~\cite{Martin:2009iq,Lai:2010vv,NNPDF:2014otw}. When going from pp to pA collisions, there have been two approaches to deal with the nuclear modification~\cite{Accardi:2012qut}, especially at the kinematic region where one probes the small-$x$ parton physics. One is a DGLAP-based approach, while the other one is the saturation-based or color glass condensate (CGC) approach. In the DGLAP-based approach, one replaces the usual proton PDFs with the nuclear modified PDFs (nPDFs)~\cite{Eskola:2013aya,Hirai:2007sx,deFlorian:2003qf,Eskola:2021nhw,Helenius:2021tof,Shen:2021eir} and follows the exact same collinear factorization. In this approach, the nuclear modification is included in the parameterization of the initial conditions for the DGLAP evolution of the nPDFs. On the other hand, in the saturation/CGC approach, gluon mergers and interactions dynamically lead to the nonlinear BK-JIMWLK evolution equations~\cite{Balitsky:1995ub,Kovchegov:1999yj,Kovchegov:1999ua,Jalilian-Marian:1997jhx,Jalilian-Marian:1998tzv,Jalilian-Marian:1997ubg,Iancu:2000hn}. For the theoretical formalism of the dijet production in the CGC framework, see for example Refs.~\cite{Marquet:2007vb,Kotko:2015ura}. See also other work~\cite{Ke:2023xeo,Ru:2019qvz,Ru:2023ars,Arleo:2020rbm,Kang:2012am} along this direction.

\begin{figure}
  \centering
\begin{tikzpicture}
    \def\R{2.8}
    \def\M{3.9}
    \coordinate (O) at (0,0);
    \coordinate (J1) at (-90:0.9*\R); 
    \coordinate (J2) at (75:0.90*\R); 
    \coordinate (M1) at (90:0.8*\M); 
    \coordinate (M2) at (0:0.35*\M); 
    
    \jetcone{O}{J2}{1.}{-0.1}
    \node[vector,below=2,right=1] at (J2) {$j_2$};

    \draw[ptmiss] (O) -- (M1) node[left=10,below=0] {$y$};
    \draw[ptmiss] (O) -- (M2) node[right=0,below=1] {$x$};
    
    \jetcone{O}{J1}{1.}{0.1}
    \node[vector,left] at (J1) {$j_1$};
  
    \draw[line width=0.6,mygreen,decorate,decoration={brace,amplitude=4}]
    ($(M1)+(94:-.21*\M)$) -- ($(J2)+(60:-0.1*\R)$)  node[midway,right=3,above=2] {$\delta\phi$};
    
  \end{tikzpicture}
  \caption{Definition of the azimuthal angular $\delta \phi$ of dijet pair production in the $x$-$y$ plane, where the transverse momentum of the leading jet $j_1$ is chosen to be aligned with the $-y$ direction for convenience. } \label{fig:dijet}
\end{figure}
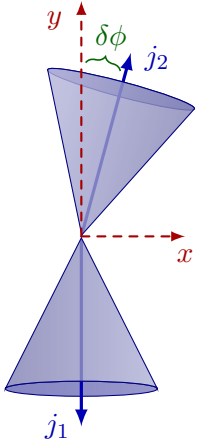

Alternatively, when one studies more differential dijet observables, e.g. dijet azimuthal decorrelation, the conventional pQCD collinear factorization could be impaired. In the nearly back-to-back region where $\delta\phi = \pi - \Delta\phi \to 0$, the perturbative expansion of the azimuthal angle decorrelation diverges due to logarithmic singularities at $\delta\phi \to 0$~\cite{Banfi:2008qs,Hautmann:2008vd}. The pioneering work in this field has highlighted the necessity of all-order resummation for accurately describing hadronic radiation, leading to a TMD-like factorization as shown below. This conclusion has been supported by numerous studies that have performed all-order resummation for various processes~\cite{Banfi:2003jj,Sun:2014gfa,Sun:2015doa,Chen:2018fqu,Sun:2018icb,Liu:2018trl,Buffing:2018ggv,Chien:2019gyf,Liu:2020dct,Liu:2020jjv,Chien:2020hzh,Kang:2020xez,delCastillo:2020omr,Hatta:2020bgy,Abdulhamid:2021xtt,delCastillo:2021znl,Hatta:2021jcd,Chien:2022wiq,Bouaziz:2022tik,Yang:2022qgk,Martinez:2022dux,Ju:2022wia,Zhang:2020onw,Shao:2023zge}. In Fig.~\ref{fig:dijet} we depict this back-to-back configuration for a narrow jet radius $(R\ll 1)$, where $R$ is the radius of the jet. Fortunately, the azimuthal decorrelation of QCD jets in the nearly back-to-back region is sensitive to the intrinsic motion of the bound partons, allowing us to perform three-dimensional (3D) quantum imaging of the proton at high-energy facilities such as the Relativistic Heavy Ion Collider (RHIC) and the Large Hadron Collider (LHC). This three dimensional structure is encoded in the transverse momentum dependent parton distribution functions (TMDPDFs), which contain both collinear and transverse momentum degrees of freedom. 

While studying the nuclear modification to the inclusive dijet pseudorapidity spectrum in pA collisions, in the DGLAP-based approach, one encodes nuclear modification inside the nPDFs within the collinear factorization formalism. The natural question is how one handles the nuclear modification of the dijet production in the nearly back-to-back region when going from pp to pA collisions. As a natural generalization, we could encode nuclear modification of back-to-back dijet production inside nuclear modified TMDPDFs (nTMDPDFs) within the TMD-like factorization formalism. Following such an approach, a recent global extraction of nuclear-modified TMDPDFs has successfully described world data for semi-inclusive electron-nucleus deep inelastic scattering and Drell-Yan processes in proton-nucleus collisions in Ref.~\cite{Alrashed:2021csd}. Furthermore, an independent cross check of this analysis was performed in Ref.~\cite{Barry:2023qqh}, verifying the results of Ref.~\cite{Alrashed:2021csd}. However, the applicability of nTMDPDFs to other processes, such as dijet production, is yet to be determined. Finally, the study of QCD jet production in forward rapidity regions where one probes small-$x$ parton dynamics is crucial for investigating the phenomenon of gluon saturation or CGC. Just like nPDF vs CGC approaches, to confirm saturation effects, it is important to have a proper understanding of the impact of nTMDPDFs vs CGC approaches in the back-to-back dijet production. For recent studies that deal with the back-to-back dijet production within the CGC formalism, see for example Refs.~\cite{Mueller:2013wwa,Taels:2022tza,Caucal:2023nci}.

Experimental measurements of the azimuthal angular decorrelations in proton-proton and proton-lead (pPb) collisions at the LHC were performed in \cite{CMS:2011hzb,CMS:2014qvs}, respectively; while in \cite{CMS:2014qvs,CMS:2018jpl} the integrated dijet azimuthal angle decorrelation in the region $\Delta\phi >2\pi/3$ was measured. The first phenomenological studies of these data have been used to further constrain the nuclear modified collinear PDFs, see for instance in~\cite{Eskola:2016oht,Eskola:2021nhw,AbdulKhalek:2022fyi}, by approximating the integrated azimuthal angular decorrelations with the dijet pseudorapidity spectrum within a next-to-leading order (NLO) collinear factorization formalism. However, in the back-to-back region, which is encapsulated by $\Delta\phi >2\pi/3$, the TMD effects, such as non-perturbative corrections and resummation can also be explored. Due to the sensitivity of these data to both collinear and transverse momentum contributions, these data can serve as a window into a simultaneous extraction of both collinear and transverse momentum effects in bound nucleons inside the heavy nucleus, which has so far not been performed.

In this study, we investigate the azimuthal angular decorrelation of dijet production in proton-proton collisions using the soft-collinear effective theory (SCET) framework \cite{Bauer:2000yr,Bauer:2001ct,Bauer:2001yt,Bauer:2002nz,Beneke:2002ph}. The utilization of the SCET framework enables us to perform QCD resummation of the large logarithmic terms in the azimuthal angle and jet radius at next-to-leading logarithmic (NLL) accuracy. Additionally, we examine the effects of nuclear modification on the azimuthal angular distribution in proton-nucleus collisions through the incorporation of nTMDPDFs and comment on the implications of our formalism to measuring nTMDPDFs as well as understanding nuclear modification of both collinear and transverse motions of the partons inside the nucleus.

Two predominant approaches are typically utilized for calculating the resummation formula in azimuthal decorrelation, known as the {\it indirect} \cite{Sun:2014gfa} and the {\it direct} \cite{Banfi:2008qs} methods. The {\it indirect} strategy focuses on the extraction of an all-order factorization and resummation formula for the two-dimensional transverse momentum imbalance $\bm q_T$ of dijet pairs and the subsequent development of the azimuthal decorrelation $\Delta\phi$ distribution originating from the $\bm q_T$ distribution. In contrast, the {\it direct} method underpins the derivation of a factorization formula for the azimuthal angular distribution in the back-to-back limit, followed by the direct computation of all-order resummation results. While the association between these two methods is explicit for Drell-Yan-like procedures, it becomes increasingly intricate for processes implicating jet production, necessitating the resummation of sizable logarithms from final-state QCD radiation. Historically, it has been demonstrated that the {\it indirect} method could induce divergences in the azimuthal integral for a narrow jet radius \cite{Buffing:2018ggv,Chien:2019gyf}. To mitigate these issues, various regularization schemes have been recommended \cite{Buffing:2018ggv,Chien:2019gyf,delCastillo:2021znl}. In this study, to evade such complexities, we have opted for the application of the {\it direct} method.

The rest of this paper is organized as follows. In section \ref{sec:fac} we first discuss the factorization and resummation formula for nearly back-to-back dijet production in proton-proton collisions. Then we present the nuclear modified resummation formula in proton-nucleus collisions. In sub-section \ref{subsec:param}, we provide information for the numerical parameterization of the non-perturbative physics as well as the non-global logarithms (NGLs). We present the numerical results using the theoretical formula, enumerate all theoretical uncertainties and compare our predictions with the LHC experimental data in sub-section \ref{subsec:results}. We also make predictions for the azimuthal decorrelation of dijet production at the LHC, as well as for the sPHENIX kinematics region at the RHIC. We summarize our paper in section \ref{sec:summary}. The details of anomalous dimensions are provided in the appendix.

\section{Factorization and resummation formula}\label{sec:fac}

In this section, we present our factorization and resummation formalism for the azimuthal decorrelation of dijet production in pp and pA collisions in the back-to-back limit.

\subsection{Factorization in SCET for pp collisions}
In the back-to-back limit and with the narrow jet approximation, the QCD modes which contribute to the dijet cross section are given by
\begin{align}
  \textbf{ hard}:&~~p_h^\mu \sim p_T (1,1,1), \\
  {\color{black} n_{a,b}\textbf{-collinear}}:&~~p_{c_i}^\mu\sim p_T\,(\delta\phi^2,1,\delta\phi)_{n_i \bar n_i}, \label{eq:pdf-mode} \\
  {\color{black} \textbf{soft}}:&~~p_s^\mu\sim p_T\,(\delta\phi,\delta\phi,\delta\phi), \label{eq:soft-mode} \\
  { \color{black} n_{c,d} \textbf{-collinear}}:&~~p_{c_i}^\mu\sim p_T\,(R^2,1,R)_{n_i \bar n_i},  \label{eq:coll-mode}\\
  { \color{black} n_{c,d} \textbf{-collinear-soft}}:&~~ p_{cs_i}^\mu\sim \frac{p_T\,\delta\phi}{R}(R^2,1,R)_{n_i \bar n_i}, \label{eq:coft-mode}
\end{align}
where the momentum $p^\mu$ is expressed in light-cone coordinates as $p^\mu \equiv (n_i\cdot p,\bar n_i\cdot p, p_{n_{i\perp}})_{n_i \bar n_i}$, and $n_i^\mu$ are light-like vectors associated with the initial-state proton beams $(n_{a,b})$ or final-state jets $(n_{c,d})$. The $n_{a,b}$-collinear, $n_{c,d}$-collinear-soft and soft modes all have the same invariant mass and will result in rapidity divergences in the factorization formula. We address these divergences using the standard Collins-Soper-Sterman (CSS) treatment \cite{Collins:1981uk,Collins:1984kg} and collinear anomaly \cite{Becher:2010tm,Becher:2011xn}  method, as explained in the next subsection. The contribution from the Glauber modes, which would result in the breaking of TMD factorization \cite{Collins:2007nk,Rogers:2010dm,Catani:2011st,Forshaw:2012bi}, is neglected in this study. The magnitude of factorization breaking effects from the Glauber mode can be explored by comparing theoretical predictions with future high-precision experimental data.

Based on the assumption of the above kinematic modes, we follow the standard steps in SCET \cite{Stewart:notes,Becher:2014oda,Schwartz:2013pla} to obtain the following factorization formula \footnote{A comprehensive description of the TMD factorization formula in the context of SCET for jet production can be found in the literature, for instance, in Refs.~\cite{Chien:2019gyf,Chien:2022wiq,delCastillo:2020omr}.}
\begin{align}\label{eq:fac}
  \frac{\mathrm{d}^4 \sigma_{\rm pp}}{\mathrm{d}y_c\,\mathrm{d}y_d\, \mathrm{d} p_T^2\,  \mathrm{d} q_x } = & \sum_{abcd} \frac{x_a x_b}{16\pi \hat s^2} \frac{1}{1+\delta_{cd}} \mathcal{C}_x\left[f_{a/p}^{\rm unsub}f_{b/p}^{\rm unsub}\, \bm{S}^{\rm unsub}_{ab\to cd, IJ}\, S^{\rm cs}_c\,S^{\rm cs}_d\right] \\
  &\times \,\bm{H}_{ab\to cd, JI}(\hat s, \hat t,\mu)\, J_c(p_T R,\mu)\, J_d(p_T R,\mu)\,, \nn
\end{align}
where we have taken the short-hand 
\begin{align}
    \mathcal{C}_x& \Big [f_{a/p}^{\rm unsub}f_{b/p}^{\rm unsub} \, \bm{S}^{\rm unsub}_{ab\to cd, IJ}\, S^{\rm cs}_c\,S^{\rm cs}_d \Big] = \int \mathrm{d}k_{ax}\,\mathrm{d}k_{bx}\,\mathrm{d}k_{cx}\,\mathrm{d}k_{dx}\,\mathrm{d}\lambda_x \, \bm{S}^{\rm unsub}_{ab\to cd, IJ}(\lambda_x,\mu,\nu)\nn \\
    & \times f_{a/p}^{\rm unsub}(x_a,k_{ax},\mu,\zeta_a/\nu^2) \, f_{b/p}^{\rm unsub}(x_b,k_{bx},\mu,\zeta_b/\nu^2)\, S^{\rm cs}_c(k_{cx},R,\mu,\nu)\,S^{\rm cs}_d(k_{dx},R,\mu,\nu) \nn \\
    &\times \delta\left(q_x-k_{ax}-k_{bx}-k_{cx}-k_{dx}-\lambda_x\right)\,.
\end{align}
The cross section is differential with respect to: the $x$ component of the transverse momentum imbalance of the jet pair ($|q_x|=p_T\delta\phi$), the outgoing rapidities of jets $c$ and $d$ ($y_{c,d}$), the jet transverse momentum ($p_T$). In this expression, $a,b,c,d$ represent parton flavors which are summed over in the cross section. The Kronecker delta symbol $\delta_{cd}$ in the prefactor on the right side of this expression arises from the symmetry factor due to identical partons in the final state. Additionally, in this expression we introduced the partonic center-of-mass energy reads $\hat s=x_a x_b s$, and $\hat t=-x_ap_T\sqrt{s}e^{-y_c}$, where $s$ is the hadronic CM energy and $x_a$ and $x_b$ represent the Bjorken variables which are defined in terms of our phase space variables through the relations
\begin{align}
  x_a = \frac{p_T}{2 E_p}\left(e^{y_c} + e^{y_d}\right),~~~  x_b = \frac{p_T}{ 2E_p} \left(e^{-y_c} + e^{-y_d}\right), 
\end{align} 
where $E_{p}$ is the energy of the incoming protons in the lab frame. 
The functions $f_{a,b/p}^{\rm unsub}$ represent the one-dimensional unsubtracted TMDPDFs for the incoming parton of flavor $a,b$ \cite{Gao:2022bzi}. For these distributions, $\mu$ and $\nu$ are standard renormalization scale and rapidity scales, while $\zeta_{a,b}$ represent the Collins-Soper parameters~\cite{Boussarie:2023izj,Collins:2011zzd}.

The function $\bm H_{ab\to cd}$ and $\bm S_{ab\to cd}$ are the hard and soft functions. In our formalism, we follow the work of \cite{Catani:1996jh,Kelley:2010fn} to organize the hard and soft functions into matrices, denoted by the bold characters. In this formalism, the IR divergent, UV finite scattering amplitudes for the 2 $\rightarrow$ 2 process can be written as vectors in color space
\begin{align}\label{eq:amp}
    \left| \mathcal{M}_{ab\rightarrow cd}\left(\hat{s},\hat{t},\hat{u},\mu,\epsilon\right)\right\rangle = \sum_I \frac{1}{\left \langle \mathcal{C}_I \mathcal{C}_I\right \rangle}\mathcal{M}_{ab\rightarrow cd}^I\left(\hat{s},\hat{t},\hat{u},\mu,\epsilon\right) \left| \mathcal{C}_I\right\rangle\,,
\end{align}
where $\left|\mathcal{C}_I\right \rangle$ denote basis vectors in the color space while $I$ is an index that runs over the dimensionality of the color space, which is determined purely through the species and the number of the external particles in the hard partonic process. The prefactors of the color basis vectors contain the kinematic contributions and the IR divergences of the amplitudes. Following the work of \cite{Kelley:2010fn}, the basis vectors are absorbed into the soft sector. We now note that the integration of the virtual partons in the amplitudes of Eq.~\eqref{eq:amp} contain interactions at the hard scale as well as interactions at scales associated with the IR modes in Eqs.~\eqref{eq:pdf-mode}, \eqref{eq:soft-mode}, \eqref{eq:coll-mode}, and \eqref{eq:coft-mode}. To define a purely hard scattering amplitude, one needs to subtract off the virtual loop contributions from these IR modes. As the virtual loop integrals of the IR modes are scaleless, this subtraction scheme swaps the IR divergences in the scattering amplitudes of Eq.~\eqref{eq:amp} to UV ones. Thus we can define the purely hard scattering amplitudes through the subtraction
\begin{align}\label{eq:ampH}
    \left| \mathcal{M}_{ab\rightarrow cd}^{H}\left(\hat{s},\hat{t},\hat{u},\mu,\epsilon\right)\right\rangle = \left| \mathcal{M}_{ab\rightarrow cd}\left(\hat{s},\hat{t},\hat{u},\mu,\epsilon\right)\right\rangle - \sum_{i} \left| \mathcal{M}_{ab\rightarrow cd}^{i}\left(\hat{s},\hat{t},\hat{u},\mu,\epsilon\right)\right\rangle\,,
\end{align}
where $i$ runs over the IR modes in \eqref{eq:pdf-mode}, \eqref{eq:soft-mode}, \eqref{eq:coll-mode} and \eqref{eq:coft-mode}. The divergences entering into the hard scattering amplitude are now UV and can therefore be handled in a multiplicative renormalization procedure. Thus we can define UV subtracted amplitudes as
\begin{align}\label{eq:ampUV}
    \left| \mathcal{M}_{ab\rightarrow cd}^{H\, \rm sub}\left(\hat{s},\hat{t},\hat{u},\mu\right)\right\rangle = \bm{Z}_H\left(\hat{s},\hat{t},\hat{u},\mu,\epsilon\right)\, \left| \mathcal{M}_{ab\rightarrow cd}^{H}\left(\hat{s},\hat{t},\hat{u},\mu,\epsilon\right)\right\rangle\,,
\end{align}
where $\bm{Z}_H$ is the hard multiplicative renormalization factor and is a matrix in color space. From this expression, the evolution of the subtracted scattering amplitudes is given by the expression
\begin{align}
    \frac{\partial}{\partial \ln{\mu}}\left| \mathcal{M}_{ab\rightarrow cd}^{H\, \rm sub}\left(\hat{s},\hat{t},\hat{u},\mu\right)\right\rangle = \bm{\Gamma}_H\left(\hat{s},\hat{t},\hat{u},\mu\right) \left| \mathcal{M}_{ab\rightarrow cd}^{H\, \rm sub}\left(\hat{s},\hat{t},\hat{u},\mu\right)\right\rangle
\end{align}
where the hard anomalous dimension is defined as
\begin{align}
    \bm{\Gamma}_H\left(\hat{s},\hat{t},\hat{u},\mu\right)=\left[\frac{\partial}{\partial \ln{\mu}} \bm{Z}_H\left(\hat{s},\hat{t},\hat{u},\mu\right) \right] \bm{Z}^{-1}_H\left(\hat{s},\hat{t},\hat{u},\mu\right)\,.
\end{align}
In the following section, we will provide the hard anomalous dimension matrix, while we will further summarize the formalism in this section. 

In SCET, the soft contributions enter as vacuum matrix elements. In our formalism, we define a $b$-space unsubtracted global soft function as
\begin{align}\label{eq:soft-new}
 \tilde{S}^{\rm unsub}_{ab\to cd}(\lambda_x,\mu,\nu)=\int \frac{\mathrm{d}b}{2\pi} e^{i \lambda_x b} \big\langle 0\big|\bar{\mathbf{T}}\big[ \boldsymbol{O}_{n_a n_b n_c n_d}^\dagger(b^\mu) \big] \mathbf{T}\big[\boldsymbol{O}_{n_a n_b n_c n_d}(0)\big]\big| 0\big\rangle\,,
\end{align}
with $\boldsymbol{O}_{n_a n_b n_c n_d}(b^\mu)=[\boldsymbol{S}_{n_a} \boldsymbol{S}_{n_b}^{\dagger} \boldsymbol{S}_{n_c}^{\dagger} \boldsymbol{S}_{n_d}](x^\mu)$. In this expression, $b^\mu=(0,b,0,0)$, $n_{i}^\mu$ are the light-like vectors defined below Eq.~\eqref{eq:coft-mode}, and $\mathbf{T}$ ($\bar{\mathbf{T}}$) represents (anti-) time ordering. The soft Wilson line is given by 
\begin{align}
  \bm S_{n_i}(x)=\mathcal{P} \exp \left[i g_s \int_{-\infty}^0 d t \, n_i \cdot \bm A_s(x+t n_i )\right], 
\end{align}
where $\mathcal{P}$ denotes path ordering. We stress that, since we derive the factorization formalism in the {\it direct} method, the transverse vector $b^\mu$ points along the $x$-direction, which is perpendicular to all vectors $n_{a,b,c,d}$. This differs from the TMD soft function which was derived in \cite{Kang:2020xez}, where the TMD factorization was derived for the two-dimensional transverse momentum imbalance of dijet pairs. As a result, the operator definition of the TMD soft function in this paper is different from that in \cite{Kang:2020xez}. The soft function in Eq.~\eqref{eq:soft-new} also enters into the factorization in the transverse energy-energy correlator event shape in \cite{Gao:2019ojf}. To define the color matrix, we follow the work of Ref.~\cite{Ahrens:2010zv} to absorb the color vectors into the soft function as
\begin{align}\label{eq:soft-mat}
  \tilde{\bm{S}}^{\rm unsub}_{ab\to cd, IJ}(\lambda_x,\mu,\nu)= \left \langle \mathcal{C}_I \left|\tilde{S}^{\rm unsub}_{ab\to cd}(\lambda_x,\mu,\nu) \right|\mathcal{C}_J\right \rangle \,,
\end{align}
where the $SU(3)$ generators in the Wilson lines beyond tree level modify the color structure of the soft color matrices.

Aside from these complications associated with the hard and soft color matrices, to describe this observable, we must account for two final-state radiative effects. Firstly, in the narrow jet approximation $(R\ll 1)$, radiative corrections of the final-state partons are encoded in the jet and collinear-soft functions, $J_i$ and $\tilde S_i^{\rm cs}$. The one loop exclusive jet function is well-known, see for instance \cite{Ellis:2010rwa}, while the one-loop calculation of the collinear-soft function can be found in the appendix of \cite{Zhang:2020onw}. In addition to the standard $\epsilon$ divergences in dimensional regularization, the collinear-soft function that enters into our factorization also contains rapidity poles. We stress that these rapidity poles enter into the direct computation of the azimuthal angle decorrelation. However these poles do not enter into the collinear-soft function for the two dimension dijet transverse momentum imbalance in \cite{Kang:2020xez}. Secondly, as the observable is insensitive to radiative emissions within the jet, this observable is non-global and is thus sensitive to NGLs \cite{Dasgupta:2001sh}. Such NGLs modify the factorization structure of the jet and collinear-soft function at two loops. The full factorization formula can be obtained by introducing the multi-Wilson structure in SCET \cite{Becher:2015hka,Becher:2016mmh}. For simplicity, we do not write down the full formula in this paper, and in the resummation calculation, we use the fitting function \cite{Dasgupta:2001sh,Dasgupta:2002bw} to include their contribution at the NLL accuracy. 

After taking these effects into account, we note that the convolution in the cross section, $\mathcal{C}_x$, can be simplified by working in $b$-space, the conjugate space to $q_x$. After performing the Fourier transform, the convolutional integral can be written as
\begin{align}
    \mathcal{C}_x& \Big [f_{a/p}^{\rm unsub}f_{b/p}^{\rm unsub} \, \bm{S}^{\rm unsub}_{ab\to cd, IJ}\, S^{\rm cs}_c\,S^{\rm cs}_d \Big] = \int \frac{\mathrm{d}b}{2\pi} e^{i b p_T \delta \phi} \, \tilde{\bm{S}}^{\rm unsub}_{ab\to cd, IJ}(b,\mu,\nu)\nn \\
    & \times \tilde{f}_{a/p}^{\rm unsub}(x_a,b,\mu,\zeta_a/\nu^2) \, \tilde{f}_{b/p}^{\rm unsub}(x_b,b,\mu,\zeta_b/\nu^2)\, \tilde{S}^{\rm cs}_c(b,R,\mu,\nu)\,\tilde{S}^{\rm cs}_d(b,R,\mu,\nu)\,,
\end{align}
where the $b$-space functions are defined as
\begin{align}
    \tilde{f}_{a/p}^{\rm unsub}(x_a,b,\mu,\zeta_a/\nu^2) = \int \mathrm{d}k_{ax}\,e^{-i k_{ax} b} \, \tilde{f}_{a/p}^{\rm unsub}(x_a,k_{ax},\mu,\zeta_a/\nu^2)\,,
\end{align}
\begin{align}
  \tilde{\bm{S}}^{\rm unsub}_{ab\to cd,IJ}(b,\mu,\nu)=\int \mathrm{d}\lambda_x\, e^{-i \lambda_x b}\, \tilde{\bm{S}}^{\rm unsub}_{ab\to cd,IJ}(\lambda_x,\mu,\nu)\,,
\end{align}
\begin{align}
    \tilde{S}^{\rm cs}_c(b,R,\mu,\nu) = \int \mathrm{d}k_{cx}\, e^{-i k_{cx} b}\,  \tilde{S}^{\rm cs}_c(k_{cx},R,\mu,\nu)\,.
\end{align}
After taking into consideration the simplification when working in $b$-space, the expression for the factorized cross section is given by the expression
\begin{align}\label{eq:fac-bspace}
    \frac{\mathrm{d}^4 \sigma_{\rm pp}}{\mathrm{d}y_c\,\mathrm{d}y_d\, \mathrm{d} p_T^2\,  \mathrm{d} q_x } = & \sum_{abcd} \frac{x_a x_b}{16\pi \hat s^2} \frac{1}{1+\delta_{cd}} \,\bm{H}_{ab\to cd, JI}(\hat s, \hat t,\mu)\, J_c(p_T R,\mu)\, J_d(p_T R,\mu) \\
    &\times \int \frac{\mathrm{d}b}{2\pi} e^{i b p_T \delta \phi} \, \tilde{\bm{S}}^{\rm unsub}_{ab\to cd, IJ}(b,\mu,\nu)\, \tilde{S}^{\rm cs}_c(b,R,\mu,\nu)\,\tilde{S}^{\rm cs}_d(b,R,\mu,\nu)\nn \\
    & \times \tilde{f}_{a/p}^{\rm unsub}(x_a,b,\mu,\zeta_a/\nu^2) \, \tilde{f}_{b/p}^{\rm unsub}(x_b,b,\mu,\zeta_b/\nu^2)\,. \nn
\end{align} 
In the following sections, we will summarize the expressions for the evolution and resummation of each contribution in this cross section.

\subsection{RG evolution and resummation formula}

In the above subsection, we have obtained a factorization formula for azimuthal angular distribution in the joint back-to-back and small jet radius region. To achieve the resummation formula, one solves the RG equations for each of the ingredients in \eqref{eq:fac-bspace}. In this section, we begin by performing resummation for pp scattering and then discuss our treatment for the pA scattering.

The hard functions for all $2\to2$ processes in massless QCD are given up to next-to-next-to-leading order (NNLO) in Ref. \cite{Broggio:2014hoa}. To ensure consistency in the expressions for the hard anomalous dimensions between this study and our work, we choose to use the same color basis as this reference. Using these bases, the hard function satisfies the RG equation as
\begin{align}
  \frac{\mathrm{d}}{\mathrm{d} \ln \mu} \boldsymbol{H}=\boldsymbol{\Gamma}_H  \boldsymbol{H}+\boldsymbol{H}  \,\boldsymbol{\Gamma}_{H}^{ \dagger}
\end{align}
where the anomalous dimension takes the form
\begin{align}\label{eq:h-adim}
     \boldsymbol{\Gamma}_{H_{ab\to cd}} & = \left[ \frac{C_H}{2}\gamma_{\rm cusp}(\alpha_s) \left(\ln\frac{\hat s}{\mu^2}-i\pi\right) + \gamma_H(\alpha_s) \right] \bm{1} + \gamma_{\rm cusp}(\alpha_s) \bm{M}_{ab\to cd},
\end{align}
with $C_H=n_q C_F + n_g C_A$ and $\gamma_H = n_q \gamma_q + n_g \gamma_g$. Here $n_q$ and $n_g$ indicate the number of quark and gluon, respectively. The matrix $\bm M$ reads
\begin{align}    
     \bm{M}_{ab\to cd} = \left(\ln r+i \pi\right) \bm{M}_{1,ab\to cd} \notag +  \ln\frac{r}{1-r} \bm{M}_{2,ab\to cd}, 
\end{align}
where the dimensionless parameter $r$ is defined as $r=-\hat t /\hat s$. The expressions for $\bm M_{1,2}$ can be found in Ref.~\cite{Broggio:2014hoa}. In this work, we consider QCD resummation at NLL accuracy, thus, we include the double logarithms anomalous dimension up to two-loop order and the single logarithms anomalous dimension up to one-loop order. The coefficients of all anomalous dimensions used in our calculation are given in the appendix \ref{app:a-dim} and we remark that the anomalous dimensions for quadrupole color and kinematic entanglement have been ignored in \eqref{eq:h-adim}, since they contribute at three-loop order and beyond \cite{Almelid:2015jia,Almelid:2017qju}. Lastly, we remark that information associated with solving the RG equations in color space is provided in \cite{Kelley:2010fn}. 

The jet functions in Eq.~\eqref{eq:fac-bspace} fulfill the RG equation
\begin{align}
  \frac{\mathrm{d}}{\mathrm{d} \ln\mu} J_i\left(p_{T} R, \mu\right)=\Gamma^{J_i}(\alpha_s) J_i\left(p_{T} R, \mu\right),
\end{align}
where the anomalous dimension of the jet is given by
\begin{align}
 \Gamma^{J_i}(\alpha_s) = - C_i \gamma_{\rm cusp}(\alpha_s) \ln \frac{p_T^2 R^2}{\mu^2} + \gamma^{J_i}(\alpha_s)\,.
\end{align}
In this expression, $C_i=C_F$ or $C_A$ is the Casimir of the parton $i$. It is worth noting that our analysis here does not account for the non-global structures in the factorization formula \eqref{eq:fac-bspace}. As shown in \cite{Chien:2019gyf}, to obtain a complete description, the contribution of non-global structures must also be incorporated. In our current study, we do not take into consideration these structures in the factorization formula. However, the leading logarithmic (LL) NGLs are resummed by a fitting function, which is explained later in the paper.

In addition to the hard and jet function, all other terms in \eqref{eq:fac-bspace} also depend on the rapidity scale $\nu$. For the TMDPDFs, we resum the large logarithms using the Collins-Soper equation. Specifically, in the Collins-11 treatment \cite{Collins:2011zzd,Boussarie:2023izj}, the properly-defined TMDPDFs are obtained by absorbing the standard TMD soft function in the Dell-Yan process, $\tilde S_{a b}(b, \mu, \nu)$, and we have
\begin{align}\label{eq:properTMDPDF}
  \tilde f_{a/p}^{\text {unsub }}\left(x_a, b, \mu, \zeta_a/\nu^2\right) \tilde f_{b/p}^{\text {unsub }}\left(x_b, b, \mu, \zeta_b/\nu^2\right) & \tilde S_{a b}(b, \mu, \nu) \\ \notag 
  &\equiv \tilde f_{a/p}\left(x_a, b, \mu,\zeta_a\right) \tilde f_{b/p}\left(x_b, b, \mu,\zeta_b\right),
\end{align}
where the rapidity divergences cancel and no explicit $\nu$-dependence in the arguments anymore. For each TMDPDF, the CSS evolution equation for the $\zeta$-dependence is given by
\begin{align}\label{eq:CSeq}
  \sqrt{\zeta_a} \frac{\mathrm{d}}{\mathrm{d}\sqrt{\zeta_a}} \tilde f_{a/p}(x_a, b,\mu,\zeta_a) = \tilde \kappa_a(b,\mu) \tilde f_{a/p}(x_a, b,\mu,\zeta_a) ,
\end{align}
where $\tilde \kappa_a(b,\mu)$ represents the Collins-Soper kernel. In the perturbative region, one has $\tilde \kappa_a(b,\mu) = -C_a \gamma_{\rm cusp}(\alpha_s)\ln \mu^2/\mu_b^2 + \mathcal{O}(\alpha_s^2)$ with $\mu_b = 2e^{-\gamma_E}/b$.  The solution reads
\begin{align}\label{eq:CSeq-solution}
  \tilde f_{a/p}(x_a, b,\mu,\zeta_{a,f}) = \tilde f_{a/p}(x_a, b,\mu,\zeta_{a,i}) \left(\sqrt{\frac{\zeta_{a,f}}{\zeta_{a,i}}}\right)^{\tilde \kappa_a(b,\,\mu) },
\end{align}
where we choose the standard Collins-Soper parameter as $\zeta_{a,i}=\zeta_{b,i}=\mu_b^2$ and $\zeta_{a,f}=\zeta_{b,f}=\hat s$. In addition, the RG equation of TMDPDFs reads 
\begin{align}
  \frac{\mathrm{d}}{\mathrm{d}\ln{\mu}}\tilde{f}_{a/p}(x_a,b,\mu,\zeta_{a,f})&=\left[C_a \gamma_{\rm cusp}(\alpha_s)\ln{\frac{\mu^2}{\zeta_{a,f}}}-2\gamma_{a}(\alpha_s)\right]\tilde{f}_{a/p}(x_a,b,\mu,\zeta_{a,f})\,,
\end{align}
where $C_a$ denote the color of the incoming parton. In comparison to the two-dimensional transverse momentum resummation formula \cite{Kang:2020xez}, the presence of rapidity divergence in the collinear-soft functions represents a new property. This divergence arises from the small jet approximation \cite{Zhang:2020onw} and requires resummation of the corresponding rapidity logarithms. Two commonly used approaches to achieve this resummation are the rapidity RG \cite{Chiu:2011qc,Chiu:2012ir} and collinear anomaly \cite{Becher:2010tm,Becher:2011xn} framework. In this study, we choose to use the collinear anomaly framework. 

In our study, we re-factorize the product of the global soft function and two collinear-soft functions. Using the collinear anomaly framework, we define a novel soft function $\bm W$ as
\begin{align}
  \bm W_{ab\to cd}(b,\mu) R^{2F_{cd}(b,\mu)}  \equiv \tilde{\bm S}_{ab\to cd}^{\rm unsub}(b,\mu,\nu)\, \tilde S^{\rm cs}_c(b,R,\mu,\nu) \, \tilde S^{\rm cs}_d(b,R,\mu,\nu) /\tilde{S}_{ab}(b,\mu,\nu),
\end{align}
where the rapidity logarithms arising from the narrow jet approximation in the collinear-soft functions are refactorized through the collinear anomaly exponent $F_{cd}=\alpha_s/(2\pi)(C_c+C_d)\ln(\mu^2/\mu_b^2) + \mathcal{O}(\alpha_s^2)$. Notice that in this expression, we have subtracted the back-to-back soft function, which has already been included in the properly-defined TMDPDFs as in Eq.~\eqref{eq:properTMDPDF}. This subtraction is required to avoid double counting of the soft modes in the final factorization formalism. Their renormalization group equations have the form as
\begin{align}
  \frac{\mathrm{d}}{\mathrm{d\ln\mu}} F_{cd}(b,\mu) & = (C_c+C_d) \gamma_{\rm cusp}(\alpha_s), \\
  \frac{\mathrm{d}}{\mathrm{d\ln\mu}} \bm W(b,\mu) & = \bm \Gamma^\dagger_W \bm W(b,\mu) +  \bm W(b,\mu) \bm \Gamma_W,
\end{align}
where $\bm \Gamma_W$ is expressed as 
\begin{align}
  \bm \Gamma_{W} =&\sum_{i<j} \mathbf{T}_i \cdot \mathbf{T}_j \gamma_{\text {cusp }}(\alpha_s) \ln \frac{ n_i \cdot n_j}{2 } \\
  & + \left[ \frac{C_c}{2}\gamma_{\rm cusp}(\alpha_s) \ln \frac{{\rm sech}^2 y_c}{4} + \frac{C_d}{2}\gamma_{\rm cusp}(\alpha_s) \ln \frac{{\rm sech}^2 y_d}{4}\right] \mathbf{1} + \mathcal{O}(\alpha_s^2), \notag
\end{align}

A rigorous test of our formalism is that we can obtain the RG invariance of the cross section as
\begin{align}
     \frac{\mathrm{d}}{\mathrm{d} \ln \mu} \mathrm{Tr}\left[\bm{H}_{ab\to cd}\bm{W}_{ab\to cd}\right] R^{2F_{cd}} \tilde{f}_{a/p} \tilde{f}_{b/p} J_c J_d = 0\,.
\end{align}
At NLL accuracy, the TMDPDF matches onto the collinear PDF through the relation
\begin{align}\label{eq:matching}
    \tilde{f}_{a/p}(x_a,b,\mu,\zeta_{a,f}) & = f_{a/p}\left(x_a,\mu_{b_*}\right) \\ 
    &\times \exp\left\{\int_{\mu_{b_*}}^\mu \frac{d\mu'}{\mu'}\left[C_a \gamma_{\rm cusp}(\alpha_s)\ln{\frac{{\mu'}^2}{\zeta_{a,f}}}-2\gamma_{a}(\alpha_s)\right]\right\}\exp \left[-S_{\mathrm{NP}}^a(b, Q_0, \sqrt{\hat{s}})\right]\,, \nn 
\end{align}
where we have used the fact that the rapidity anomalous dimension vanishes at the scale $\mu_b$, $\tilde \kappa_a(b,\mu_b) = 0$ at NLL accuracy and the $f$ on the right hand side denotes the collinear PDF. Additionally, to circumvent the issue of the Landau pole in the large $b$ limit, we have introduced the $b_*$ prescription that will be discussed in more detail in Sec.~\ref{sec:Numerics}. Lastly in Eq.~\eqref{eq:matching}, we have introduced the non-perturbative Sudakov, which parameterizes the intrinsic motion of the bound partons and depends on the initial TMD scale $Q_0$.

Combining the results for the hard, jet, TMDPDFs, and soft functions at NLL accuracy, our final resummed expression for azimuthal angular distribution is 
\begin{align}
\label{eq:res}
  & \frac{\mathrm{d}^4  \sigma_{pp}}{\mathrm{d}y_c\, \mathrm{d}y_d\, \mathrm{d} p_T^2\, \mathrm{d} \delta \phi } =  \sum_{abcd} \frac{p_T}{16\pi \hat s^2} \frac{1}{1+\delta_{cd}} \int_0^\infty \frac{2\mathrm{d}b}{\pi} \cos( b p_T \delta\phi) x_a \tilde{f}_{a/p}(x_a,\mu_{b_*}) x_b  \tilde{f}_{b/p}(x_b,\mu_{b_*}) \notag \\
& \times \exp \left\{-\int_{\mu_{b_*}}^{\mu_h} \frac{\mathrm{d} \mu}{\mu}\left[\gamma_{\text {cusp}}\left(\alpha_s\right) C_H \ln \frac{\hat{s}}{\mu^2}+2 \gamma_H\left(\alpha_s\right)\right]\right\} \notag \\
& \times \sum_{K K^{\prime}} \exp \left[-\int_{\mu_{b_*}}^{\mu_h} \frac{\mathrm{d} \mu}{\mu} \gamma_{\text {cusp}}\left(\alpha_s\right)\left(\lambda_K+\lambda_{K^{\prime}}^*\right)\right] H_{K K^{\prime}}\left(\hat s, \hat t, \mu_h\right) W_{K^{\prime} K}\left(b_*, \mu_{b_*}\right)  \notag \\
& \times \exp \left[-\int_{\mu_{b_*}}^{\mu_j} \frac{\mathrm{d} \mu}{\mu} \Gamma^{J_c}\left(\alpha_s\right) -\int_{\mu_{b_*}}^{\mu_j} \frac{\mathrm{d} \mu}{\mu} \Gamma^{J_d}\left(\alpha_s\right)\right] U_{\mathrm{NG}}^c\left(\mu_{b_*}, \mu_j\right) U_{\mathrm{NG}}^d\left(\mu_{b_*}, \mu_j\right) \notag \\
& \times \exp \left[-S_{\mathrm{NP}}^a(b, Q_0, \sqrt{\hat{s}})-S_{\mathrm{NP}}^{b}(b, Q_0, \sqrt{\hat{s}})\right]\,.
\end{align}
In this expression, the quantity $\lambda_K$ represents the eigenvalue of the matrices $\bm M_{1,2}$. In the small jet radius regime, the resummation of NGLs is achieved through a non-linear RG evolution between the jet and collinear-soft functions \cite{Chien:2019gyf} that is contained in the $U_{\rm NG}^i$ functions. Lastly, $\mu_h$ and $\mu_j$ are the hard and jet scales which will be discussed in Sec.~\ref{subsec:param}.

\subsection{Nuclear modified formalism for pA collisions}
Having established the factorization and resummation for dijet production in pp collisions in the previous section, in this section we extend this formalism to incorporate the nuclear modifications in pA collisions. 

As we mentioned in the Introduction, for observables that can be described by the collinear factorization formalism, a DGLAP-based approach can be used to deal with the nuclear modification when going from pp to pA collisions. In this approach, one assumes the same collinear factorization while replacing the proton PDFs with the nuclear modified PDFs~\cite{Eskola:2013aya,Hirai:2007sx,deFlorian:2003qf,Eskola:2021nhw,Helenius:2021tof}. Now for the azimuthal decorrelation of dijet production in the nearly back-to-back region,  a TMD factorization and resummation in Eq.~\eqref{eq:fac} is derived. Thus, as a natural generalization of the idea implemented in nPDFs, when going from pp to pA collisions, we assume that the same factorization and resummation formalism in Eq.~\eqref{eq:fac} holds for pA collisions, while replacing the proton TMDPDF $\tilde{f}_{b/p}$ with the nuclear modified TMDPDF $\tilde{f}_{b/A}$ for the target nucleus. The nTMDPDF $\tilde{f}_{b/A}(x_b,b,\mu,\zeta_{b,f})$ contains the nuclear modification of both collinear (associated with $x$) and transverse (associated with $b$) motions for the partons inside the nucleus. Follow the assumptions made in Ref.~\cite{Alrashed:2021csd}, these nuclear modification will be absorbed into the non-perturbative parameterizations for the collinear PDF and the non-perturbative Sudakov. Thus under this assumption the nTMDPDF $\tilde{f}_{b/A}(x_b,b,\mu,\zeta_{b,f})$ can be matched onto the nPDF through the NLL relation
\begin{align}
\label{eq:nTMD}
    \tilde{f}_{b/A}(x_b,b,\mu,\zeta_{b,f}) & = f_{b/A}\left(x_b,\mu_{b_*}\right) \\ 
    &\times \exp\left\{\int_{\mu_{b_*}}^\mu \frac{d\mu'}{\mu'}\left[C_b\gamma_{\rm cusp}(\alpha_s)\ln{\frac{{\mu'}^2}{\zeta_{b,f}}}-2\gamma_{b}(\alpha_s)\right]\right\}\exp \left[-S_{\mathrm{NP}}^{b,A}(b, Q_0, \sqrt{\hat{s}})\right]\,. \nn 
\end{align}
Here, besides the collinear nPDF $f_{b/A}\left(x_b,\mu_{b_*}\right)$, 
we have introduced the medium modified non-perturbative Sudakov $S_{\mathrm{NP}}^{b,A}(b, Q_0, \sqrt{\hat{s}})$, whose parameterization will be discussed in the next section. Note that we only keep the leading power term in the OPE matching in Eq.~\eqref{eq:nTMD} where the nTMDPDF is matched onto the collinear nPDF. In principle, there could be power corrections ${\mathcal O}\left(b^2Q_s^2(A)\right)$ in the expansion which are associated with higher-twist nuclear matrix elements~\cite{Kang:2013wca}. Here $Q_s(A)$ is a dynamical scale, often referred to as the saturation scale~\cite{Gelis:2010nm}, associated with multiple scattering in the nuclear medium. We do not consider the effect of such power corrections in this paper. 

With this replacement for nTMDPDF and following the same resummation procedure, the factorization and resummation formalism for back-to-back dijet production at NLL accuracy in pA collisions is given by
\begin{align}
\label{eq:reseA}
  & \frac{\mathrm{d}^4  \sigma_{pA}}{\mathrm{d}y_c\, \mathrm{d}y_d\, \mathrm{d} p_T^2\, \mathrm{d} \delta \phi } =  \sum_{abcd} \frac{p_T}{16\pi \hat s^2} \frac{1}{1+\delta_{cd}} \int_0^\infty \frac{2\mathrm{d}b}{\pi} \cos( b p_T \delta\phi) x_a \tilde{f}_{a/p}(x_a,\mu_{b_*}) x_b  \tilde{f}_{b/A}(x_b,\mu_{b_*}) \notag \\
& \times \exp \left\{-\int_{\mu_{b_*}}^{\mu_h} \frac{\mathrm{d} \mu}{\mu}\left[\gamma_{\text {cusp}}\left(\alpha_s\right) C_H \ln \frac{\hat{s}}{\mu^2}+2 \gamma_H\left(\alpha_s\right)\right]\right\} \notag \\
& \times \sum_{K K^{\prime}} \exp \left[-\int_{\mu_{b_*}}^{\mu_h} \frac{\mathrm{d} \mu}{\mu} \gamma_{\text {cusp}}\left(\alpha_s\right)\left(\lambda_K+\lambda_{K^{\prime}}^*\right)\right] H_{K K^{\prime}}\left(\hat s, \hat t, \mu_h\right) W_{K^{\prime} K}\left(b_*, \mu_{b_*}\right)  \notag \\
& \times \exp \left[-\int_{\mu_{b_*}}^{\mu_j} \frac{\mathrm{d} \mu}{\mu} \Gamma^{J_c}\left(\alpha_s\right) -\int_{\mu_{b_*}}^{\mu_j} \frac{\mathrm{d} \mu}{\mu} \Gamma^{J_d}\left(\alpha_s\right)\right] U_{\mathrm{NG}}^c\left(\mu_{b_*}, \mu_j\right) U_{\mathrm{NG}}^d\left(\mu_{b_*}, \mu_j\right) \notag \\
& \times \exp \left[-S_{\mathrm{NP}}^a(b, Q_0, \sqrt{\hat{s}})-S_{\mathrm{NP}}^{b,A}(b, Q_0, \sqrt{\hat{s}})\right]\,.
\end{align}
In the next section, we will discuss our parameterization for both the collinear nPDFs $\tilde{f}_{b/A}(x_b,\mu_{b_*})$ and our nuclear modified Sudakov factor $S_{\mathrm{NP}}^{b,A}(b, Q_0, \sqrt{\hat{s}})$.

\section{Numerical parameterization and results}\label{sec:Numerics}
\subsection{Parameterization}
\label{subsec:param}
To capture the resummation of the NGLs, we follow the prescription of Ref.~\cite{Dasgupta:2001sh} to parameterize the $U$ function as
\begin{align}
  U_{\mathrm{NG}}^i\left(\mu_{b_*}, \mu_{j}\right)=\exp \left[-C_{i} C_A \frac{\pi^{2}}{3} u^{2} \frac{1+(a u)^{2}}{1+(b u)^{c}}\right],
\end{align}
where $u=\ln[\alpha_s(\mu_{b_*})/\alpha_s(\mu_j)]/\beta_0$, $a=0.85 \, C_A$, $b=0.86\,C_A$ and $c=1.33$ \cite{Dasgupta:2001sh}. Since the factorized formula \eqref{eq:fac} involves two jet functions, the square of $U_{\rm NG}$ is required to incorporate the NGL resummation associated with each jet. 

In previous work on CSS resummation, the $b_*$-prescription was introduced along with non-perturbative Sudakov factors, which were modeled through various functional forms and obtained by fitting to experimental data \cite{Collins:2014jpa,Aidala:2014hva,Sun:2014dqm,Landry:2002ix,Konychev:2005iy,Bacchetta:2017gcc,Bacchetta:2022awv}. In this work, we follow the standard $b_*$-prescription where
\begin{align}\label{eq:bstar}
  b_* \equiv b /\sqrt{1+b^2/b_{\rm max}^2}\,,
  \qquad
  \mu_{b_*} = 2 e^{-\gamma_E}/b_*\,,
\end{align}
as in \cite{Collins:1984kg}. Since we also need to study the impact of the nuclear modification on the azimuthal angular distribution in proton-nucleus collisions, we adopt the same functional form used in Refs. \cite{Sun:2014dqm,Echevarria:2020hpy} which was employed in the extraction of nTMDPDFs \cite{Alrashed:2021csd}. Specifically, the non-perturbative Sudakov factors in the last line of Eq. \eqref{eq:res} are given by
\begin{align}
  S_{\mathrm{NP}}^{a, b}(b, Q_0, Q)=g_1^f b^2+\frac{g_2}{2} \frac{C_{a, b}}{C_F} \ln \frac{Q}{Q_0} \ln \frac{b}{b_*}, 
  \label{eq:Sudakov}
\end{align}
with $g_1^f=0.106\,\mathrm{GeV}^{-2}$, $g_2=0.84$ and $Q_0^2=2.4 \, \mathrm{GeV}^2$. Finally, in our numerical calculations the intrinsic scales in the resummation formula \eqref{eq:res} are chosen as
\begin{align}\label{eq:hj_scale}
    \mu_h = p_T,~~~\mu_j = p_T R.
\end{align}

To obtain numerical results for the pA collisions, we need a parameterization for nTMDPDF in Eq.~\eqref{eq:nTMD}, which contains both the collinear and transverse motion for partons inside the nucleus. To describe the medium modifications to the collinear PDF, specifically $f_{b/A}\left(x_b,\mu_{b_*}\right)$, we follow the parameterization in \cite{Alrashed:2021csd} to use the EPPS16 parameterization given in \cite{Eskola:2016oht} while describing the collinear PDF for the proton, we use CT14nlo parameterization \cite{Dulat:2015mca}. On the other hand, for the nuclear modification to the transverse motion in nTMDPDFs, we follow the parameterization of Ref.~\cite{Alrashed:2021csd} to have a nuclear modified Sudakov factor $S_{\mathrm{NP}}^{b,A}(b, Q_0, \sqrt{\hat{s}})$. Specifically, we replace the $g_1$ parameter in Eq.~\eqref{eq:Sudakov}, which accounts for the broadening effects of transverse momentum within the nucleus. Adopting the functional form obtained from the global extraction in \cite{Alrashed:2021csd}, we take
\begin{align}
\label{eq:aN}
g_1^A = g_1^f + a_N L, \quad\textrm{with}\quad a_N=0.016\, \textrm{GeV}^{-2}~\textrm{and}~ L=A^{1/3}-1,
\end{align}
where $g_1^A$ characterizes the transverse momentum width of partons inside the nucleus and is also proportional to the saturation scale in the small-$x$ region \cite{Mueller:2016xoc}. Thus the nuclear modified non-perturbative Sudakov factor is defined as 
\begin{align}
\label{eq:nSudakov}
    S_{\mathrm{NP}}^{b,A}(b, Q_0, Q)=g_1^A b^2+\frac{g_2}{2} \frac{C_{b}}{C_F} \ln \frac{Q}{Q_0} \ln \frac{b}{b_*}.
\end{align}

\subsection{Numerical results}
\label{subsec:results}
In this section, we present our numerical results for the pp and pA resummation formulas derived in the previous section. Specifically, we apply the theory formalisms in Eqs.~\eqref{eq:res} and \eqref{eq:reseA} for pp and pA collisions, respectively and compare them with the existing experimental data. We discuss applications of this formalism to measuring nuclear modifications to collinear and transverse motions of partons in nTMDPDFs. We also provide predictions for the dijet production in the forward rapidity region in pPb collisions at the LHC, as well as in pAu collisions for the sPHENIX kinematics at the RHIC.

\begin{figure}[t]
  \centering
  \includegraphics[width=0.48\textwidth,clip]{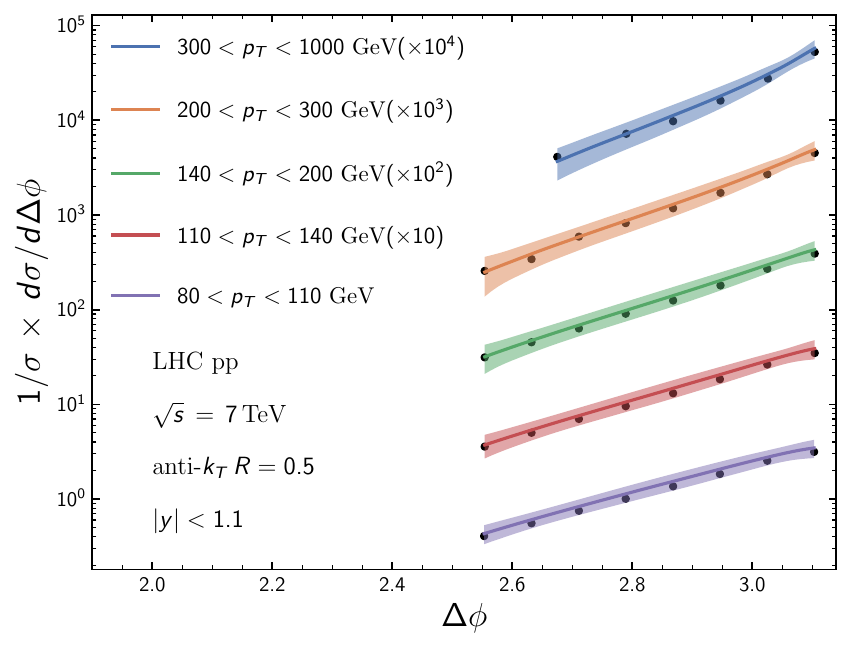}
  \includegraphics[width=0.49\textwidth,clip]{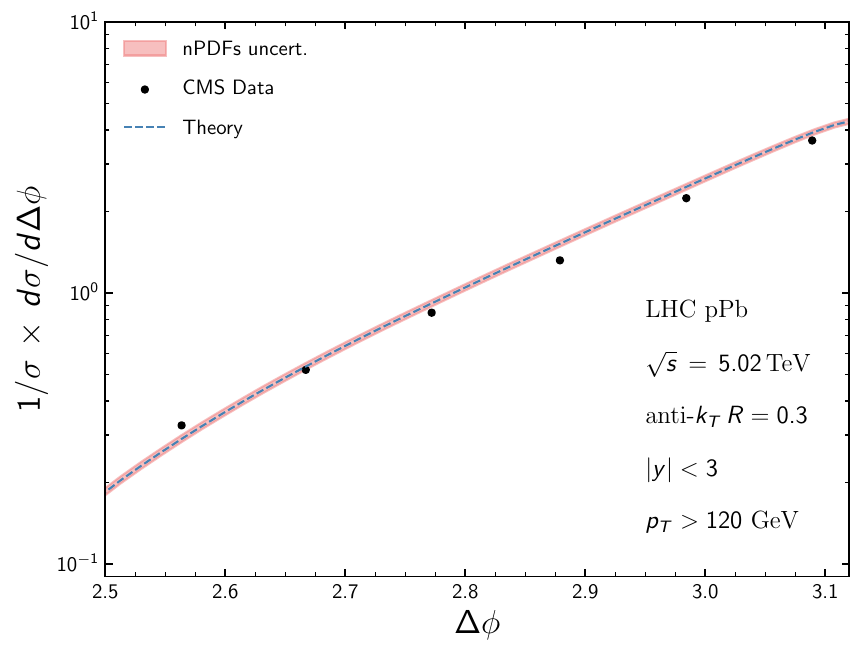}
  \caption{Left: Comparison between theoretical calculations of the azimuthal decorrelation with the CMS data \cite{CMS:2011hzb}, where $\Delta\phi$ is the difference in the azimuthal angle between two leading jets. The solid curves are the theoretical distributions, which are normalized by dividing the LO cross section.  The black dots are the CMS results, and the uncertainties of the data are smaller than the symbol size used in the plot. The colored bands indicate theoretical uncertainties from the variation of hard and jet scales. Right: A comparison of the dijet azimuthal angle decorrelation in pPb collisions from the CMS collaboration at the LHC~\cite{CMS:2014qvs}.}\label{fig:theory-data}
\end{figure}

A comprehensive investigation into the QCD resummation of azimuthal decorrelation in dijet production in pp collisions was carried out in \cite{Sun:2014gfa} using the {\it indirect }method, as outlined in the introduction. The analysis successfully resummed the large logarithmic terms of azimuthal angle and jet radius at NLL and LL accuracy, respectively, while ignoring the contribution from NGLs. In our work, we present a resummation formula for azimuthal decorrelation in the {\it direct} method. This approach accounts for both the large logarithmic terms of the azimuthal angle and jet radius at NLL accuracy, including the contribution from NGLs. As a verification of the formula, we compare its theoretical predictions to measurements of dijet production in proton-proton collisions taken by the CMS collaboration at the LHC with $\sqrt{s}=7$ TeV, as presented in the left panel of Fig.~\ref{fig:theory-data}. The QCD jets were reconstructed using the anti-$k_T$ algorithm \cite{Cacciari:2008gp} with a radius of $R=0.5$ and the rapidities of each jet were limited to $|y_{c,d}|<1.1$. Additionally, to construct the denominator of the normalized $\Delta\phi$ distribution, we use the LO expression for the cross section. The data, shown as black dots, covers five bins ranging from $80$ GeV to $1$ TeV for the jet transverse momentum $p_T$. The theoretical results, displayed as lines of different colors, are found to agree well with the measurements in the back-to-back region across all $p_T$ bins. Besides, we also show the uncertainties from scale variations, which are given by the colored bands. Here we vary the hard and jet scales by a factor of two around their default values as defined in Eq. \eqref{eq:hj_scale}, and the total uncertainty bands are obtained by the envelope of all the variations. Since the non-perturbative Sudakov factor in Eq. \eqref{eq:Sudakov} is fitted at the canonical scale $\mu_{b_*}$, we do not include uncertainties from its variations. It is noteworthy that the contribution of Glauber modes, which can potentially violate TMD factorization, is not considered in this analysis. Therefore, the magnitude of naive factorization breaking due to Glauber modes can be evaluated by comparing theoretical predictions with future high-precision experimental measurements. 

On the right side of Fig.~\ref{fig:theory-data}, we plot the azimuthal angle decorrelation in pPb collisions at $\sqrt{s} = 5.02$ TeV from the CMS collaboration~\cite{CMS:2014qvs}. The data is integrated within the region $|y_{c,d}|<3$ and the jets were reconstructed using an anti-$k_T$ algorithm with $R = 0.3$. In the theory calculation, we implement nTMDPDFs which encode the nuclear modification to the collinear (as in nPDFs $f_{b/A}(x, \mu_{b_*}$)) and transverse motion (as in the broadening parameter $a_N$) in Eqs.~\eqref{eq:nTMD} and \eqref{eq:nSudakov}. The dashed blue theory curve is computed with the central fit of nPDFs in the EPPS16 parametrization and the broadening parameter $a_N$ in Eq.~\eqref{eq:aN}. The red band is the uncertainty from the nPDFs fit. Our calculations agree with the experimental data in the back-to-back region $\Delta \phi\sim \pi$. We also observed in both plots of Fig.~\ref{fig:theory-data} that our theoretical prediction starts to deviate from the experimental data points away from the back-to-back region, i.e. when $\Delta\phi$ moves away from $\pi$. This is expected since our formalism applies only to the resummation region. Such a discrepancy can be corrected by including the fixed order calculation for the dijet azimuthal angular decorrelation, see for instance \cite{Sun:2015doa}. 

\begin{figure}[t]
  \centering
  \includegraphics[width=1\textwidth,clip]{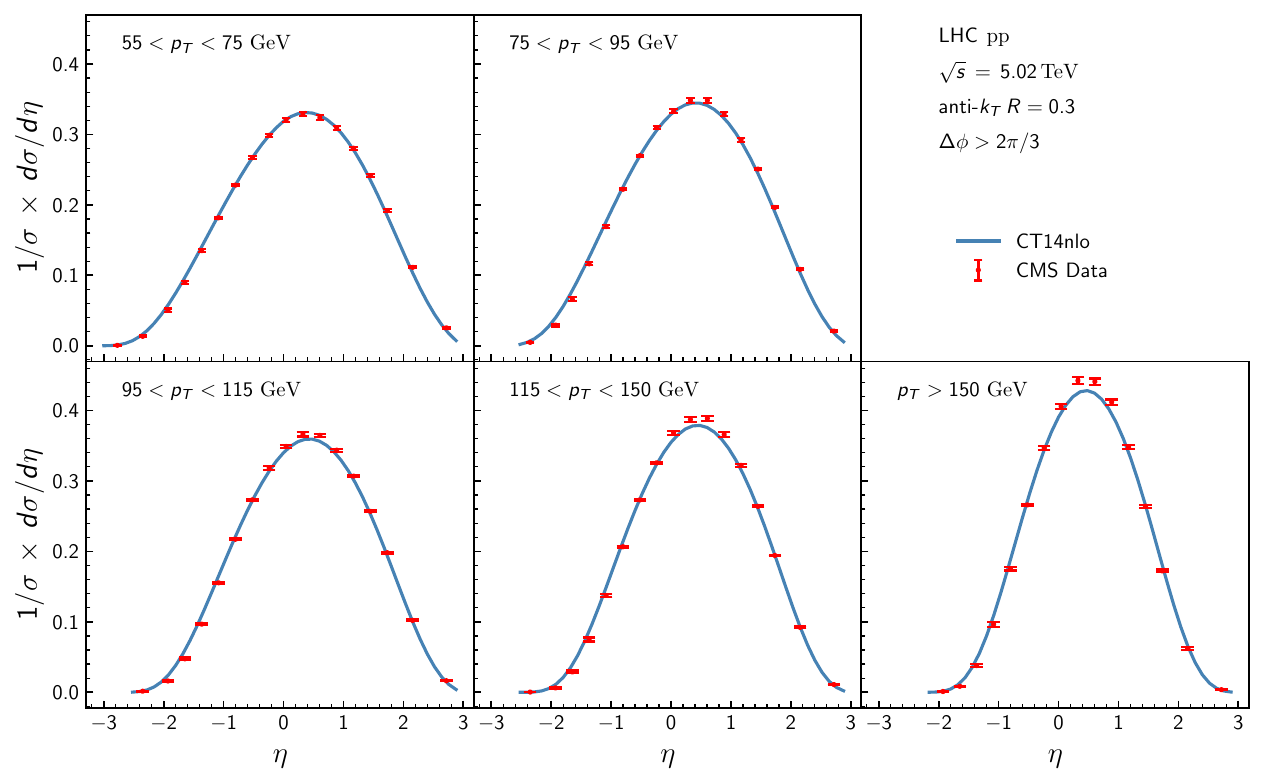}
  \caption{Theoretical calculations for the dijet integrated angular decorrelation plotted as a function of the pseudorapidity $\eta$ are compared with the CMS data \cite{CMS:2018jpl} in proton-proton collisions for different kinematic cuts. The spectra were shifted by +0.465 to match the dijet pseudorapidity $\eta$ range of the corresponding proton-lead collisions. }\label{fig:CMS_pp_ETA}
\end{figure}

\begin{figure}[hbt!]
  \centering
  \includegraphics[width=1\textwidth,clip]{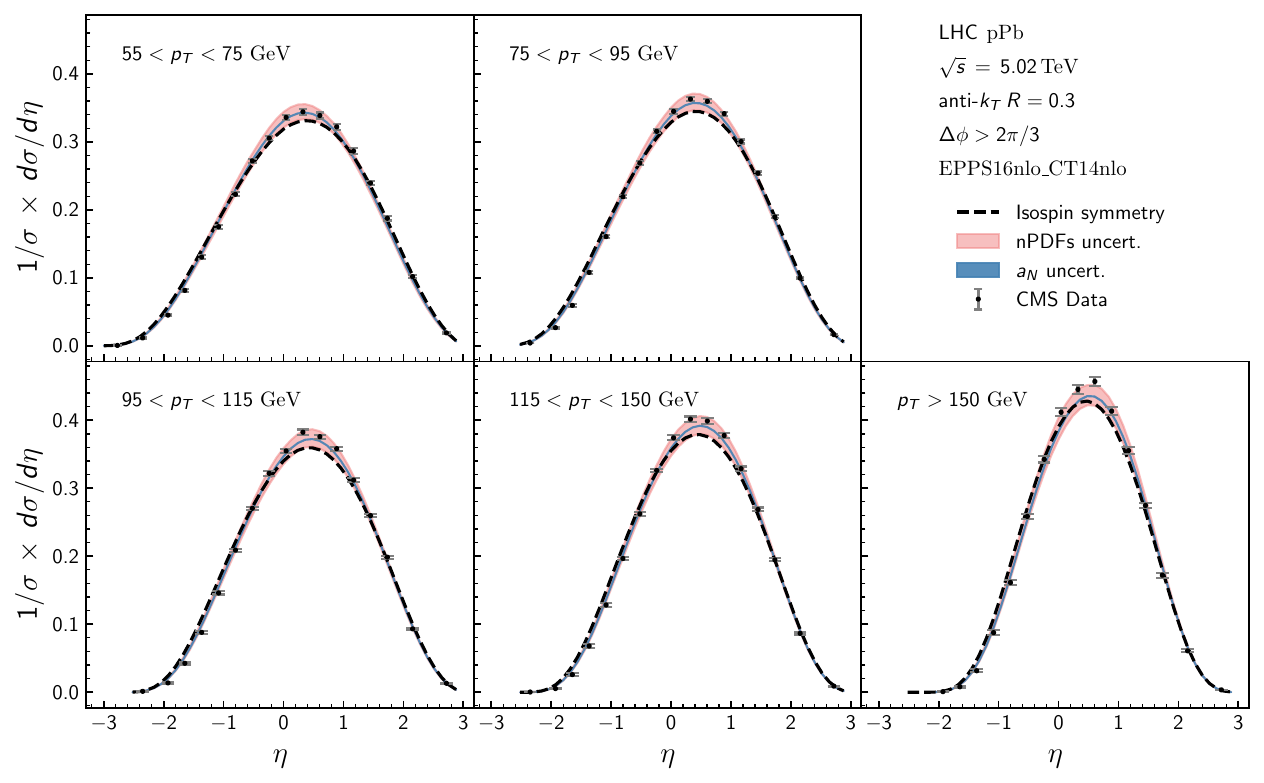}
  \caption{Theoretical calculations for dijet integrated angular decorrelation plotted as a function of the pseudorapidity $\eta$ are compared with the CMS data \cite{CMS:2018jpl} in proton-lead collisions for different kinematic cuts. }\label{fig:CMS_PA_ETA}
\end{figure}

\begin{figure}[hbt!]
  \centering
  \includegraphics[width=1\textwidth,clip]{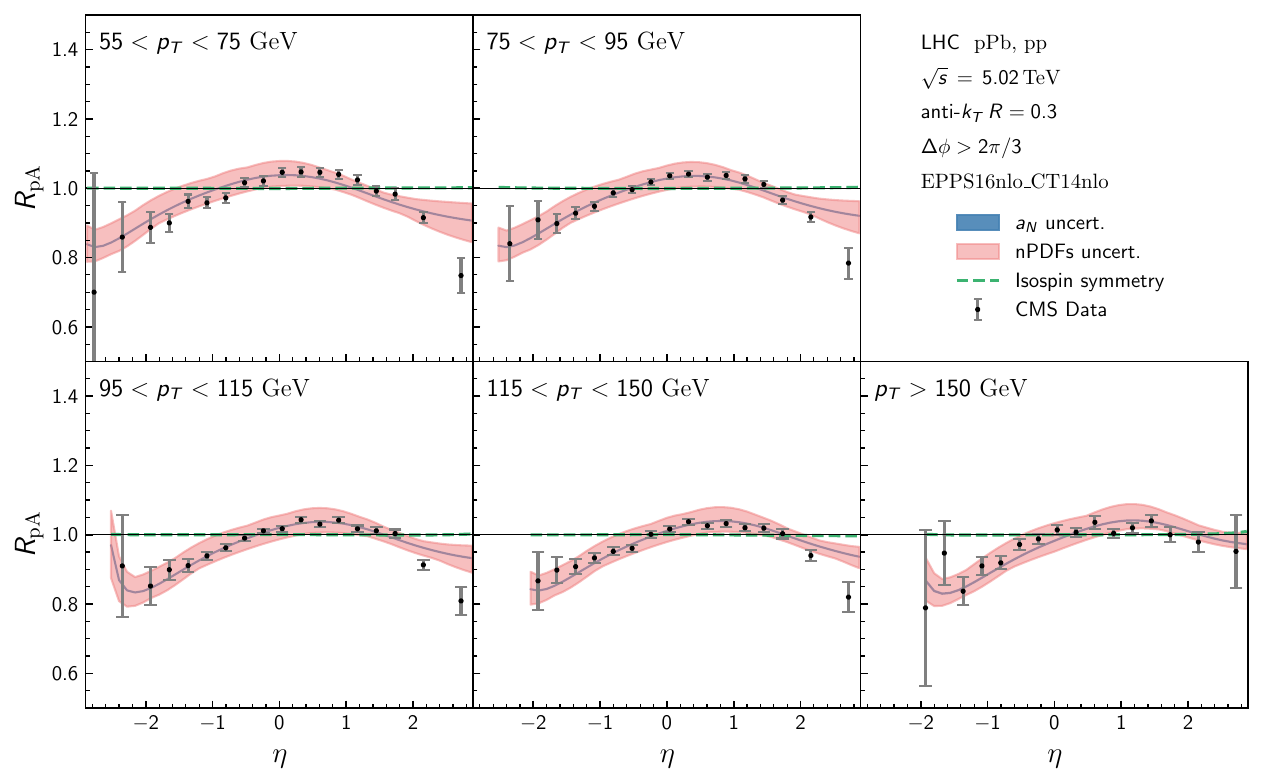}
  \caption{Theoretical calculations for the nuclear modification factor $R_{\rm pA}$ plotted as a function of the pseudorapidity $\eta$ are compared with the CMS data \cite{CMS:2018jpl} for different jet transverse momentum cuts.  }\label{fig:CMS_PAtopp_ETA}
\end{figure}

In Fig.~\ref{fig:CMS_pp_ETA}, we present a comparison between the NLL pQCD calculations of the dijet integrated angular decorrelation plotted as a function of the dijet pseudorapidity $\eta = (y_c+y_d)/2$ in pp collisions, respectively, and corresponding experimental measurement taken by CMS \cite{CMS:2018jpl}. The data are categorized based on the transverse momentum ($p_T$) of the dijet system where the jet radius is $R=0.3$. To enable an extensive comparison of the two datasets, the experimental measurements are superimposed onto the theoretical predictions, allowing us to evaluate the compatibility between the model and the experimental data. In the theoretical calculation, we integrate $\Delta\phi$ from $2\pi/3$ to $\pi$ using \eqref{eq:res} and to form the $\sigma$ in the denominator of the integrated azimuthal angle decorrelation, we integrate over the pseudorapidity coverage of both jets following the experimental cuts. We observe that our theory calculations describes the experimental data quite well. 

In Fig.~\ref{fig:CMS_PA_ETA}, we present our NLL calculation of the integrated angular decorrelation plotted as a function of the pseudorapidity in pPb collisions and the CMS experimental data in \cite{CMS:2018jpl}. To demonstrate the importance of nuclear modification to parton dynamics in the nucleus, we include a calculation where one only takes into account the isospin effect. In other words, going from pp to pA collisions, one only replaces the PDFs in the proton by the PDFs that include the isospin effect
\begin{align}
    f_{i/A}\left(x,\mu\right) = \frac{Z}{A} f_{i/p}\left(x,\mu\right) +\frac{A-Z}{A}f_{i/n}\left(x,\mu\right)\,,
\end{align}
where $Z$ is the atomic number of the nucleus while $f_{i/p}$ and $f_{i/n}$ denote the PDFs of the proton and neutron. We find that the calculations with the isospin effect alone undershoots the data rather significantly, especially in the mid-rapidity region $0 \lesssim \eta \lesssim 1$ where an antishadowing effect is evident from the data~\cite{Eskola:2016oht,Eskola:2021nhw}. On the other hand, the central blue theory curve is computed with the central fit of nPDFs in the EPPS16 parametrization and the broadening parameter $a_N$ in Eq.~\eqref{eq:aN}. We further considered the uncertainty band associated with the collinear nPDFs as well as the broadening parameter $a_N$. It is evident that our formalism with nuclear modification implemented in nTMDPDFs describe the CMS pPb collision data well though the size of the uncertainties from the broadening parameter $a_N$ is very small. This behavior is expected as the $a_N$ parameter acts to broaden the intrinsic width of the partons. At the large $p_T$ values of the CMS data, this broadening is small compared to the large transverse momentum that is generated from the resummation. Experimental data at smaller values of $p_T$, which should be measurable at RHIC, will then depend more strongly on this parameter. However, the small dependence on $a_N$ indicates that both the integrated and unintegrated azimuthal angle decorrelation can be used to measure the collinear contribution to the nTMDPDFs.

To quantify the nuclear modification, we adopt the usual definition for the nuclear modification factor
\begin{align}
  R_\textrm{pA} = \left.\frac{1}{A}\frac{\mathrm{d}^4  \sigma_{pA}}{\mathrm{d}y_c\, \mathrm{d}y_d\, \mathrm{d} p_T^2\, \mathrm{d} \Delta \phi }\right/\frac{\mathrm{d}^4  \sigma_{pp}}{\mathrm{d}y_c\, \mathrm{d}y_d\, \mathrm{d} p_T^2\, \mathrm{d} \Delta \phi }.
\end{align}
In Fig.~\ref{fig:CMS_PAtopp_ETA}, we present the nuclear modification factor $R_{\rm pA}$ as a function of dijet rapidity $\eta$ between our theory calculations and corresponding experimental data taken by the CMS collaboration at the LHC~\cite{CMS:2018jpl}. In this plot, we have included a central curve as well as considered the uncertainty band associated with the nPDF and the broadening parameter $a_N$ in nTMDPDFs. We have also included a prediction taking into account the isospin effect alone. The nuclear modification factor $R_{\rm pA}$ with the isospin effect alone is almost unity as indicated by the dashed green curve. This is because the dijet production at this energy is mostly sensitive to the gluon distribution inside the nucleus and thus the isospin symmetry applied to $u$ and $d$ flavors does not play an important role here. On the other hand, we observe a strong consistency between the central curve of the NLL pQCD prediction with nTMDPDFs and the experimental data. However, we find that our calculations do not describe the strong suppression in the CMS data in the proton's forward region where $\eta\gtrsim 2$ and the probed parton momentum fraction $x\sim 10^{-2}$ inside the nucleus. Since this modification in our nTMDPDFs formalism is mainly driven by the collinear nPDFs in the EPPS16 parametrization, as commented in~\cite{Eskola:2016oht,Eskola:2021nhw}, this remains an open question. As our formalism neglects all final-state interactions associated with Glauber interaction with the jets~\cite{Idilbi:2008vm,Ovanesyan:2011xy,Rothstein:2016bsq}, we suspect that the cause of these discrepancies lies in these final-state effects~\cite{Kang:2017frl}. Addressing this discrepancy is vitally important for understanding the gluon distribution of the bound nucleons at this relatively small $x$ region.

\begin{figure}[t]
  \centering
  \includegraphics[width=0.328\textwidth,clip]{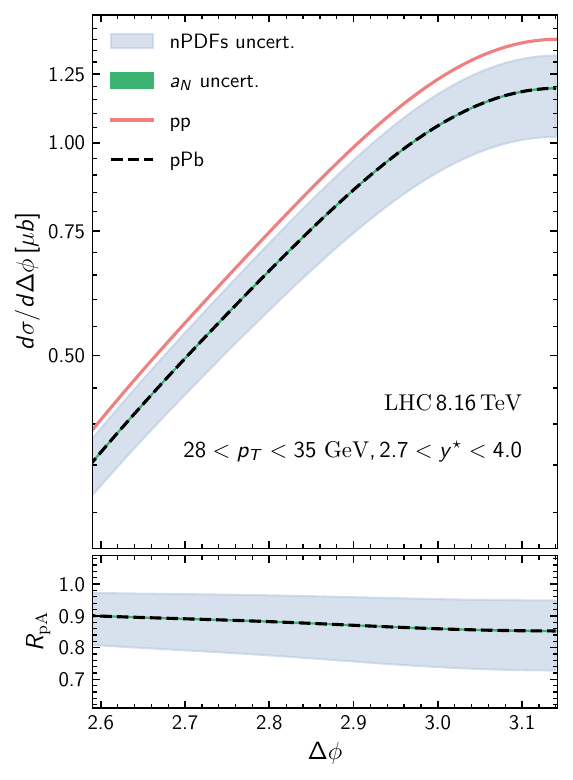}
    \includegraphics[width=0.32\textwidth,clip]{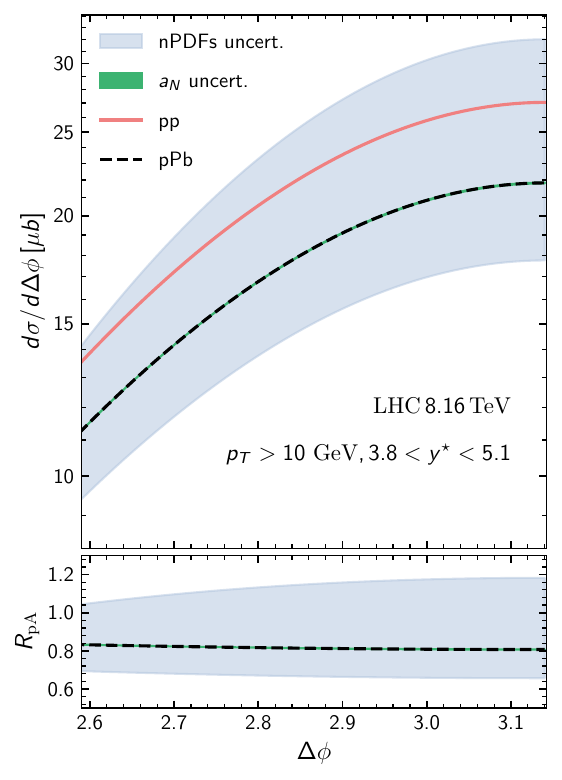}
    \includegraphics[width=0.32\textwidth,clip]{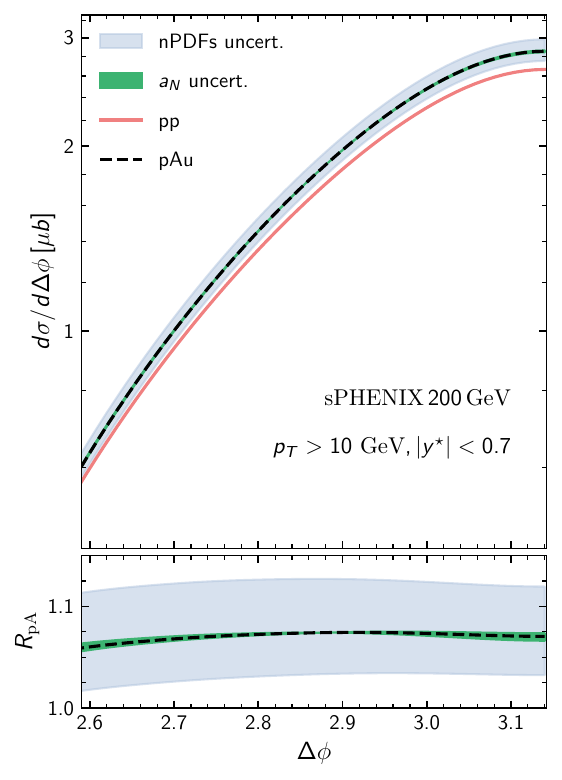}
  \caption{Top: The azimuthal angular distribution in pp (red curve) and pA (black curve) collisions for ATLAS (Left), ALICE (Middle), and sPHENIX (Right). In the lower panel, we plot the nuclear modification factor $R_\textrm{pA}$.}\label{fig:LHC_pp_PA}
\end{figure}

In the left and middle panels of Fig.~\ref{fig:LHC_pp_PA} we present the results of our calculation for the azimuthal angular distribution in pp and pA collisions in forward rapidity regions at the ATLAS and ALICE kinematics at the LHC. In the right panel, we present the results of the decorrelation for the sPHENIX kinematics at the RHIC. In our study, we adopt the same kinematic cuts at the LHC as used in Ref.~\cite{Al-Mashad:2022zbq} and at the RHIC in Ref.~\cite{Belmont:2023fau}, which are defined as follows:
\begin{enumerate}
  \item $28\,\textrm{GeV}<p_T<35 \,\textrm{GeV}$ and $2.7<y^*_{c,d}<4.0$ for the FCal calorimeter of the ATLAS at the LHC\,, 
  \item $p_T>10 \,\textrm{GeV}$ and $3.8<y^*_{c,d}<5.1$ for the upgraded FoCal of the ALICE at the LHC\,,
  \item $p_T>10\,\textrm{GeV}$ and $|y^*_{c,d}|<0.7$ for sPHENIX at the RHIC\,,
\end{enumerate}
where $y^\ast$ denotes the jet rapidity in both the pp and pA center of mass frame. The upper panels in Fig.~\ref{fig:LHC_pp_PA} show the azimuthal angular distributions in proton-proton (red curves) and proton-nucleus (black curves) collisions, while the lower panels display the nuclear modification factor $R_\textrm{pA}$. Our results indicate that in the back-to-back region, the suppression from the nTMDPDFs is substantial, with a reduction of around $20\%$ for the ATLAS and $30\%$ for the ALICE kinematics, similar to the nuclear modification reported in Ref.~\cite{Al-Mashad:2022zbq} where a saturation-based formalism is used. This is due to the shadowing effect in the small-$x$ region where the probed $x\sim 10^{-4}$. On the other hand, our calculation predicts a small enhancement $\sim 5\%$ for the sPHENIX kinematics because of the anti-shadowing effect at $x\sim 0.1$ probed in the sPHENIX experiment. Once again in the left two panels, we see that the size of the broadening parameter $a_N$ is small in comparison to the uncertainty of the nPDFs. This behavior is once again expected as the LHC produces jets with large values of $p_T$. In the right panel, we see that the size of the uncertainty from the broadening parameter $a_N$ grows larger, indicating that lower $p_T$ jets serve as a better probe of the transverse dynamics of the bound nucleons.

\section{Summary}\label{sec:summary}

In this paper, we derived a new resummation formula for the azimuthal decorrelation in dijet production in proton-proton collisions using SCET. By utilizing the {\it direct} method, we were able to account for both large logarithmic terms of the azimuthal angle and jet radius. We compared our theoretical predictions with experimental data from the CMS collaboration and found a strong agreement. We further proposed an approach to deal with the nuclear modification for nearly back-to-back dijet production in proton-nucleus collisions by introducing nuclear modified transverse momentum dependent parton distribution functions (nTMDPDFs). The nTMDPDFs contain nuclear modification to both the collinear and transverse motions for the partons inside the nucleus. Following a simple model for nTMDPDFs in our previous work that encodes nuclear modification to collinear dynamics in collinear nPDFs while nuclear modification to transverse motion in a broadening parameter, we present theoretical calculations for dijet production in proton-nucleus collisions and show good agreement with the existing experimental data at the LHC. Additionally, we presented our results for the forward rapidity region at the LHC and for the mid-rapidity region for sPHENIX at the RHIC. We applied the formula to two kinematic cuts relevant to the FCal calorimeter of the ATLAS and the upgraded FoCal of the ALICE. The results showed significant suppression of about $20\%$ for the ATLAS and $30\%$ for the ALICE in the back-to-back limit, due to the shadowing effect in the small-$x\sim 10^{-4}$ region. This suppression is of the same order as previous results within the saturation-based model. On the other hand, our calculation predicts a small enhancement $\sim 5\%$ for the sPHENIX kinematics because of the anti-shadowing effect with $x\sim 0.1$ probed in the sPHENIX experiment. Overall, this study represents an important step towards a more complete understanding of azimuthal decorrelation in dijet production and the role of nuclear modification effects. In future work, we see important applications of our formalism, e.g. in performing a simultaneous fit to both collinear and transverse momentum dependent contributions to the transverse momentum dependent distributions in nuclei. It would also be interesting to extend our results to other kinematic regions and incorporate the contributions from higher-order corrections, as well as to generalize our formalism to describe dijet production in the polarized scattering~\cite{STAR:2023xvk}. 

\acknowledgments
M.G. and D.Y.S. are supported by the National Science Foundations of China under Grant No.~12275052 and No.~12147101 and the Shanghai Natural Science Foundation under Grant No.~21ZR1406100. Z.K. is supported by the National Science Foundation under grant No.~PHY-1945471.  J.T. is supported by the Department of Energy at LANL through the LANL/LDRD Program under project number~20220715PRD1. C.Z.~is supported by the National Science Foundations of China under Grant No. 12147125, No.~12275052 and No.~12147101 and the Shanghai Natural Science Foundation under Grant No.~21ZR1406100.

\appendix

\section{Anomalous dimension}\label{app:a-dim}
The QCD $\beta$-function and the cusp and non-cusp anomalous dimensions are expanded as
\begin{equation}
\beta(\alpha_{s})=-2\alpha_{s}\sum_{n=0}^{\infty}\beta_{n}\left(\frac{\alpha_{s}}{4\pi}\right)^{n+1},\quad\gamma(\alpha_s)=\sum_{n=0}^{\infty}\gamma_{n}\left(\frac{\alpha_{s}}{4\pi}\right)^{n+1}.
\end{equation}
The two-loop coefficients of the $\beta$-function and the cusp anomalous
dimensions, and the one-loop coefficient of the non-cusp anomalous
dimensions read,
\begin{align}
&\beta_{0}=\frac{11}{3}C_{A}-\frac{4}{3}T_{F}n_{f},\quad\beta_{1}=\frac{34}{3}C_{A}^{2}-\frac{20}{3}T_{F}C_{A}n_{f}-4T_{F}C_{F}n_{f}, \notag \\
&\gamma_{0}^{\text{cusp }}=4,\quad\gamma_{1}^{\text{cusp }}=\left(\frac{268}{9}-\frac{4\pi^{2}}{3}\right)C_{A}-\frac{80}{9}T_{F}n_{f}, \notag \\
&\gamma^{J_i}_0 = -2\gamma_0^i, \quad \gamma_0^q = -3C_F,\quad \gamma_0^g=-\beta_0,
\end{align}
with $T_{F}=1/2,\:C_{A}=3,\:C_{F}=4/3,\:n_{f}=5.$

\bibliographystyle{JHEP}
\bibliography{jet.bib}

\providecommand{\href}[2]{#2}\begingroup\raggedright\begin{thebibliography}{100}

\bibitem{Accardi:2012qut}
A.~Accardi et~al., \emph{{Electron Ion Collider: The Next QCD Frontier}:
  {Understanding the glue that binds us all}},
  \href{https://doi.org/10.1140/epja/i2016-16268-9}{\emph{Eur. Phys. J. A}
  {\bfseries 52} (2016) 268},
  [\href{https://arxiv.org/abs/1212.1701}{{\ttfamily 1212.1701}}].

\bibitem{Albacete:2013ei}
J.~L. Albacete et~al., \emph{{Predictions for $p+$Pb Collisions at
  $\sqrt{s_{NN}} = 5\, {\rm TeV}$ }},
  \href{https://doi.org/10.1142/S0218301313300075}{\emph{Int. J. Mod. Phys. E}
  {\bfseries 22} (2013) 1330007},
  [\href{https://arxiv.org/abs/1301.3395}{{\ttfamily 1301.3395}}].

\bibitem{Aschenauer:2016our}
E.-C. Aschenauer et~al., \emph{{The RHIC Cold QCD Plan for 2017 to 2023: A
  Portal to the EIC}},  \href{https://arxiv.org/abs/1602.03922}{{\ttfamily
  1602.03922}}.

\bibitem{Belmont:2023fau}
R.~Belmont et~al., \emph{{Predictions for the sPHENIX physics program}},  in
  \emph{{RBRC Workshop:~Predictions for sPHENIX}}, 5, 2023,
  \href{https://arxiv.org/abs/2305.15491}{{\ttfamily 2305.15491}}.

\bibitem{Collins:1989gx}
J.~C. Collins, D.~E. Soper and G.~F. Sterman, \emph{{Factorization of Hard
  Processes in QCD}},
  \href{https://doi.org/10.1142/9789814503266_0001}{\emph{Adv. Ser. Direct.
  High Energy Phys.} {\bfseries 5} (1989) 1--91},
  [\href{https://arxiv.org/abs/hep-ph/0409313}{{\ttfamily hep-ph/0409313}}].

\bibitem{Martin:2009iq}
A.~D. Martin, W.~J. Stirling, R.~S. Thorne and G.~Watt, \emph{{Parton
  distributions for the LHC}},
  \href{https://doi.org/10.1140/epjc/s10052-009-1072-5}{\emph{Eur. Phys. J. C}
  {\bfseries 63} (2009) 189--285},
  [\href{https://arxiv.org/abs/0901.0002}{{\ttfamily 0901.0002}}].

\bibitem{Lai:2010vv}
H.-L. Lai, M.~Guzzi, J.~Huston, Z.~Li, P.~M. Nadolsky, J.~Pumplin et~al.,
  \emph{{New parton distributions for collider physics}},
  \href{https://doi.org/10.1103/PhysRevD.82.074024}{\emph{Phys. Rev. D}
  {\bfseries 82} (2010) 074024},
  [\href{https://arxiv.org/abs/1007.2241}{{\ttfamily 1007.2241}}].

\bibitem{NNPDF:2014otw}
{\scshape NNPDF} collaboration, R.~D. Ball et~al., \emph{{Parton distributions
  for the LHC Run II}},
  \href{https://doi.org/10.1007/JHEP04(2015)040}{\emph{JHEP} {\bfseries 04}
  (2015) 040}, [\href{https://arxiv.org/abs/1410.8849}{{\ttfamily 1410.8849}}].

\bibitem{Eskola:2013aya}
K.~J. Eskola, H.~Paukkunen and C.~A. Salgado, \emph{{A perturbative QCD study
  of dijets in p+Pb collisions at the LHC}},
  \href{https://doi.org/10.1007/JHEP10(2013)213}{\emph{JHEP} {\bfseries 10}
  (2013) 213}, [\href{https://arxiv.org/abs/1308.6733}{{\ttfamily 1308.6733}}].

\bibitem{Hirai:2007sx}
M.~Hirai, S.~Kumano and T.~H. Nagai, \emph{{Determination of nuclear parton
  distribution functions and their uncertainties in next-to-leading order}},
  \href{https://doi.org/10.1103/PhysRevC.76.065207}{\emph{Phys. Rev. C}
  {\bfseries 76} (2007) 065207},
  [\href{https://arxiv.org/abs/0709.3038}{{\ttfamily 0709.3038}}].

\bibitem{deFlorian:2003qf}
D.~de~Florian and R.~Sassot, \emph{{Nuclear parton distributions at
  next-to-leading order}},
  \href{https://doi.org/10.1103/PhysRevD.69.074028}{\emph{Phys. Rev. D}
  {\bfseries 69} (2004) 074028},
  [\href{https://arxiv.org/abs/hep-ph/0311227}{{\ttfamily hep-ph/0311227}}].

\bibitem{Eskola:2021nhw}
K.~J. Eskola, P.~Paakkinen, H.~Paukkunen and C.~A. Salgado, \emph{{EPPS21: a
  global QCD analysis of nuclear PDFs}},
  \href{https://doi.org/10.1140/epjc/s10052-022-10359-0}{\emph{Eur. Phys. J. C}
  {\bfseries 82} (2022) 413},
  [\href{https://arxiv.org/abs/2112.12462}{{\ttfamily 2112.12462}}].

\bibitem{Helenius:2021tof}
I.~Helenius, M.~Walt and W.~Vogelsang, \emph{{NNLO nuclear parton distribution
  functions with electroweak-boson production data from the LHC}},
  \href{https://doi.org/10.1103/PhysRevD.105.094031}{\emph{Phys. Rev. D}
  {\bfseries 105} (2022) 094031},
  [\href{https://arxiv.org/abs/2112.11904}{{\ttfamily 2112.11904}}].

\bibitem{Shen:2021eir}
S.~Shen, P.~Ru and B.-W. Zhang, \emph{{Imaging nuclear modifications on parton
  distributions with triple-differential dijet cross sections in proton-nucleus
  collisions}}, \href{https://doi.org/10.1103/PhysRevD.105.096025}{\emph{Phys.
  Rev. D} {\bfseries 105} (2022) 096025},
  [\href{https://arxiv.org/abs/2112.11819}{{\ttfamily 2112.11819}}].

\bibitem{Balitsky:1995ub}
I.~Balitsky, \emph{{Operator expansion for high-energy scattering}},
  \href{https://doi.org/10.1016/0550-3213(95)00638-9}{\emph{Nucl. Phys. B}
  {\bfseries 463} (1996) 99--160},
  [\href{https://arxiv.org/abs/hep-ph/9509348}{{\ttfamily hep-ph/9509348}}].

\bibitem{Kovchegov:1999yj}
Y.~V. Kovchegov, \emph{{Small$-x$ $F_2$ structure function of a nucleus
  including multiple pomeron exchanges}},
  \href{https://doi.org/10.1103/PhysRevD.60.034008}{\emph{Phys. Rev. D}
  {\bfseries 60} (1999) 034008},
  [\href{https://arxiv.org/abs/hep-ph/9901281}{{\ttfamily hep-ph/9901281}}].

\bibitem{Kovchegov:1999ua}
Y.~V. Kovchegov, \emph{{Unitarization of the BFKL pomeron on a nucleus}},
  \href{https://doi.org/10.1103/PhysRevD.61.074018}{\emph{Phys. Rev. D}
  {\bfseries 61} (2000) 074018},
  [\href{https://arxiv.org/abs/hep-ph/9905214}{{\ttfamily hep-ph/9905214}}].

\bibitem{Jalilian-Marian:1997jhx}
J.~Jalilian-Marian, A.~Kovner, A.~Leonidov and H.~Weigert, \emph{{The Wilson
  renormalization group for low x physics: Towards the high density regime}},
  \href{https://doi.org/10.1103/PhysRevD.59.014014}{\emph{Phys. Rev. D}
  {\bfseries 59} (1998) 014014},
  [\href{https://arxiv.org/abs/hep-ph/9706377}{{\ttfamily hep-ph/9706377}}].

\bibitem{Jalilian-Marian:1998tzv}
J.~Jalilian-Marian, A.~Kovner, A.~Leonidov and H.~Weigert, \emph{{Unitarization
  of gluon distribution in the doubly logarithmic regime at high density}},
  \href{https://doi.org/10.1103/PhysRevD.59.034007}{\emph{Phys. Rev. D}
  {\bfseries 59} (1999) 034007},
  [\href{https://arxiv.org/abs/hep-ph/9807462}{{\ttfamily hep-ph/9807462}}].

\bibitem{Jalilian-Marian:1997ubg}
J.~Jalilian-Marian, A.~Kovner and H.~Weigert, \emph{{The Wilson renormalization
  group for low x physics: Gluon evolution at finite parton density}},
  \href{https://doi.org/10.1103/PhysRevD.59.014015}{\emph{Phys. Rev. D}
  {\bfseries 59} (1998) 014015},
  [\href{https://arxiv.org/abs/hep-ph/9709432}{{\ttfamily hep-ph/9709432}}].

\bibitem{Iancu:2000hn}
E.~Iancu, A.~Leonidov and L.~D. McLerran, \emph{{Nonlinear gluon evolution in
  the color glass condensate. 1.}},
  \href{https://doi.org/10.1016/S0375-9474(01)00642-X}{\emph{Nucl. Phys. A}
  {\bfseries 692} (2001) 583--645},
  [\href{https://arxiv.org/abs/hep-ph/0011241}{{\ttfamily hep-ph/0011241}}].

\bibitem{Marquet:2007vb}
C.~Marquet, \emph{{Forward inclusive dijet production and azimuthal
  correlations in pA collisions}},
  \href{https://doi.org/10.1016/j.nuclphysa.2007.09.001}{\emph{Nucl. Phys. A}
  {\bfseries 796} (2007) 41--60},
  [\href{https://arxiv.org/abs/0708.0231}{{\ttfamily 0708.0231}}].

\bibitem{Kotko:2015ura}
P.~Kotko, K.~Kutak, C.~Marquet, E.~Petreska, S.~Sapeta and A.~van Hameren,
  \emph{{Improved TMD factorization for forward dijet production in
  dilute-dense hadronic collisions}},
  \href{https://doi.org/10.1007/JHEP09(2015)106}{\emph{JHEP} {\bfseries 09}
  (2015) 106}, [\href{https://arxiv.org/abs/1503.03421}{{\ttfamily
  1503.03421}}].

\bibitem{Ke:2023xeo}
W.~Ke, Y.-Y. Zhang, H.~Xing and X.-N. Wang, \emph{{eHIJING: an Event Generator
  for Jet Tomography in Electron-Ion Collisions}},
  \href{https://arxiv.org/abs/2304.10779}{{\ttfamily 2304.10779}}.

\bibitem{Ru:2019qvz}
P.~Ru, Z.-B. Kang, E.~Wang, H.~Xing and B.-W. Zhang, \emph{{Global extraction
  of the jet transport coefficient in cold nuclear matter}},
  \href{https://doi.org/10.1103/PhysRevD.103.L031901}{\emph{Phys. Rev. D}
  {\bfseries 103} (2021) L031901},
  [\href{https://arxiv.org/abs/1907.11808}{{\ttfamily 1907.11808}}].

\bibitem{Ru:2023ars}
P.~Ru, Z.-B. Kang, E.~Wang, H.~Xing and B.-W. Zhang, \emph{{Probing the jet
  transport coefficient of cold nuclear matter in electron-ion collisions}},
  \href{https://arxiv.org/abs/2302.02329}{{\ttfamily 2302.02329}}.

\bibitem{Arleo:2020rbm}
F.~Arleo and C.-J. Na\"\i{}m, \emph{{Nuclear p$_{\perp}$-broadening of
  Drell-Yan and quarkonium production from SPS to LHC}},
  \href{https://doi.org/10.1007/JHEP07(2020)220}{\emph{JHEP} {\bfseries 07}
  (2020) 220}, [\href{https://arxiv.org/abs/2004.07188}{{\ttfamily
  2004.07188}}].

\bibitem{Kang:2012am}
Z.-B. Kang and J.-W. Qiu, \emph{{Nuclear modification of vector boson
  production in proton-lead collisions at the LHC}},
  \href{https://doi.org/10.1016/j.physletb.2013.03.030}{\emph{Phys. Lett. B}
  {\bfseries 721} (2013) 277--283},
  [\href{https://arxiv.org/abs/1212.6541}{{\ttfamily 1212.6541}}].

\bibitem{Banfi:2008qs}
A.~Banfi, M.~Dasgupta and Y.~Delenda, \emph{{Azimuthal decorrelations between
  QCD jets at all orders}},
  \href{https://doi.org/10.1016/j.physletb.2008.05.065}{\emph{Phys. Lett. B}
  {\bfseries 665} (2008) 86--91},
  [\href{https://arxiv.org/abs/0804.3786}{{\ttfamily 0804.3786}}].

\bibitem{Hautmann:2008vd}
F.~Hautmann and H.~Jung, \emph{{Angular correlations in multi-jet final states
  from $k_\perp$-dependent parton showers}},
  \href{https://doi.org/10.1088/1126-6708/2008/10/113}{\emph{JHEP} {\bfseries
  10} (2008) 113}, [\href{https://arxiv.org/abs/0805.1049}{{\ttfamily
  0805.1049}}].

\bibitem{Banfi:2003jj}
A.~Banfi and M.~Dasgupta, \emph{{Dijet rates with symmetric $E_t$ cuts}},
  \href{https://doi.org/10.1088/1126-6708/2004/01/027}{\emph{JHEP} {\bfseries
  01} (2004) 027}, [\href{https://arxiv.org/abs/hep-ph/0312108}{{\ttfamily
  hep-ph/0312108}}].

\bibitem{Sun:2014gfa}
P.~Sun, C.~P. Yuan and F.~Yuan, \emph{{Soft Gluon Resummations in Dijet
  Azimuthal Angular Correlations in Hadronic Collisions}},
  \href{https://doi.org/10.1103/PhysRevLett.113.232001}{\emph{Phys. Rev. Lett.}
  {\bfseries 113} (2014) 232001},
  [\href{https://arxiv.org/abs/1405.1105}{{\ttfamily 1405.1105}}].

\bibitem{Sun:2015doa}
P.~Sun, C.~P. Yuan and F.~Yuan, \emph{{Transverse Momentum Resummation for
  Dijet Correlation in Hadronic Collisions}},
  \href{https://doi.org/10.1103/PhysRevD.92.094007}{\emph{Phys. Rev. D}
  {\bfseries 92} (2015) 094007},
  [\href{https://arxiv.org/abs/1506.06170}{{\ttfamily 1506.06170}}].

\bibitem{Chen:2018fqu}
L.~Chen, G.-Y. Qin, L.~Wang, S.-Y. Wei, B.-W. Xiao, H.-Z. Zhang et~al.,
  \emph{{Study of Isolated-photon and Jet Momentum Imbalance in $pp$ and $PbPb$
  collisions}},
  \href{https://doi.org/10.1016/j.nuclphysb.2018.06.013}{\emph{Nucl. Phys.}
  {\bfseries B933} (2018) 306--319},
  [\href{https://arxiv.org/abs/1803.10533}{{\ttfamily 1803.10533}}].

\bibitem{Sun:2018icb}
P.~Sun, B.~Yan, C.~P. Yuan and F.~Yuan, \emph{{Resummation of High Order
  Corrections in $Z$ Boson Plus Jet Production at the LHC}},
  \href{https://arxiv.org/abs/1810.03804}{{\ttfamily 1810.03804}}.

\bibitem{Liu:2018trl}
X.~Liu, F.~Ringer, W.~Vogelsang and F.~Yuan, \emph{{Lepton-jet Correlations in
  Deep Inelastic Scattering at the Electron-Ion Collider}},
  \href{https://doi.org/10.1103/PhysRevLett.122.192003}{\emph{Phys. Rev. Lett.}
  {\bfseries 122} (2019) 192003},
  [\href{https://arxiv.org/abs/1812.08077}{{\ttfamily 1812.08077}}].

\bibitem{Buffing:2018ggv}
M.~G.~A. Buffing, Z.-B. Kang, K.~Lee and X.~Liu, \emph{{A transverse momentum
  dependent framework for back-to-back photon+jet production}},
  \href{https://arxiv.org/abs/1812.07549}{{\ttfamily 1812.07549}}.

\bibitem{Chien:2019gyf}
Y.-T. Chien, D.~Y. Shao and B.~Wu, \emph{{Resummation of Boson-Jet Correlation
  at Hadron Colliders}},
  \href{https://doi.org/10.1007/JHEP11(2019)025}{\emph{JHEP} {\bfseries 11}
  (2019) 025}, [\href{https://arxiv.org/abs/1905.01335}{{\ttfamily
  1905.01335}}].

\bibitem{Liu:2020dct}
X.~Liu, F.~Ringer, W.~Vogelsang and F.~Yuan, \emph{{Lepton-jet Correlation in
  Deep Inelastic Scattering}},
  \href{https://doi.org/10.1103/PhysRevD.102.094022}{\emph{Phys. Rev. D}
  {\bfseries 102} (2020) 094022},
  [\href{https://arxiv.org/abs/2007.12866}{{\ttfamily 2007.12866}}].

\bibitem{Liu:2020jjv}
X.~Liu, F.~Ringer, W.~Vogelsang and F.~Yuan, \emph{{Factorization and its
  Breaking in Dijet Single Transverse Spin Asymmetries in $pp$ Collisions}},
  \href{https://arxiv.org/abs/2008.03666}{{\ttfamily 2008.03666}}.

\bibitem{Chien:2020hzh}
Y.-T. Chien, R.~Rahn, S.~Schrijnder~van Velzen, D.~Y. Shao, W.~J. Waalewijn and
  B.~Wu, \emph{{Recoil-free azimuthal angle for precision boson-jet
  correlation}},
  \href{https://doi.org/10.1016/j.physletb.2021.136124}{\emph{Phys. Lett. B}
  {\bfseries 815} (2021) 136124},
  [\href{https://arxiv.org/abs/2005.12279}{{\ttfamily 2005.12279}}].

\bibitem{Kang:2020xez}
Z.-B. Kang, K.~Lee, D.~Y. Shao and J.~Terry, \emph{{The Sivers Asymmetry in
  Hadronic Dijet Production}},
  \href{https://doi.org/10.1007/JHEP02(2021)066}{\emph{JHEP} {\bfseries 02}
  (2021) 066}, [\href{https://arxiv.org/abs/2008.05470}{{\ttfamily
  2008.05470}}].

\bibitem{delCastillo:2020omr}
R.~F. del Castillo, M.~G. Echevarria, Y.~Makris and I.~Scimemi, \emph{{TMD
  factorization for dijet and heavy-meson pair in DIS}},
  \href{https://doi.org/10.1007/JHEP01(2021)088}{\emph{JHEP} {\bfseries 01}
  (2021) 088}, [\href{https://arxiv.org/abs/2008.07531}{{\ttfamily
  2008.07531}}].

\bibitem{Hatta:2020bgy}
Y.~Hatta, B.-W. Xiao, F.~Yuan and J.~Zhou, \emph{{Anisotropy in Dijet
  Production in Exclusive and Inclusive Processes}},
  \href{https://doi.org/10.1103/PhysRevLett.126.142001}{\emph{Phys. Rev. Lett.}
  {\bfseries 126} (2021) 142001},
  [\href{https://arxiv.org/abs/2010.10774}{{\ttfamily 2010.10774}}].

\bibitem{Abdulhamid:2021xtt}
M.~I. Abdulhamid et~al., \emph{{Azimuthal correlations of high transverse
  momentum jets at next-to-leading order in the parton branching method}},
  \href{https://doi.org/10.1140/epjc/s10052-022-09997-1}{\emph{Eur. Phys. J. C}
  {\bfseries 82} (2022) 36},
  [\href{https://arxiv.org/abs/2112.10465}{{\ttfamily 2112.10465}}].

\bibitem{delCastillo:2021znl}
R.~F. del Castillo, M.~G. Echevarria, Y.~Makris and I.~Scimemi,
  \emph{{Transverse momentum dependent distributions in dijet and heavy hadron
  pair production at EIC}},
  \href{https://doi.org/10.1007/JHEP03(2022)047}{\emph{JHEP} {\bfseries 03}
  (2022) 047}, [\href{https://arxiv.org/abs/2111.03703}{{\ttfamily
  2111.03703}}].

\bibitem{Hatta:2021jcd}
Y.~Hatta, B.-W. Xiao, F.~Yuan and J.~Zhou, \emph{{Azimuthal angular asymmetry
  of soft gluon radiation in jet production}},
  \href{https://doi.org/10.1103/PhysRevD.104.054037}{\emph{Phys. Rev. D}
  {\bfseries 104} (2021) 054037},
  [\href{https://arxiv.org/abs/2106.05307}{{\ttfamily 2106.05307}}].

\bibitem{Chien:2022wiq}
Y.-T. Chien, R.~Rahn, D.~Y. Shao, W.~J. Waalewijn and B.~Wu, \emph{{Precision
  boson-jet azimuthal decorrelation at hadron colliders}},
  \href{https://doi.org/10.1007/JHEP02(2023)256}{\emph{JHEP} {\bfseries 02}
  (2023) 256}, [\href{https://arxiv.org/abs/2205.05104}{{\ttfamily
  2205.05104}}].

\bibitem{Bouaziz:2022tik}
H.~Bouaziz, Y.~Delenda and K.~Khelifa-Kerfa, \emph{{Azimuthal decorrelation
  between a jet and a Z boson at hadron colliders}},
  \href{https://doi.org/10.1007/JHEP10(2022)006}{\emph{JHEP} {\bfseries 10}
  (2022) 006}, [\href{https://arxiv.org/abs/2207.10147}{{\ttfamily
  2207.10147}}].

\bibitem{Yang:2022qgk}
H.~Yang et~al., \emph{{Back-to-back azimuthal correlations in $\mathrm {Z}
  +$jet events at high transverse momentum in the TMD parton branching method
  at next-to-leading order}},
  \href{https://doi.org/10.1140/epjc/s10052-022-10715-0}{\emph{Eur. Phys. J. C}
  {\bfseries 82} (2022) 755},
  [\href{https://arxiv.org/abs/2204.01528}{{\ttfamily 2204.01528}}].

\bibitem{Martinez:2022dux}
A.~B. Martinez and F.~Hautmann, \emph{{Azimuthal di-jet correlations with
  parton branching TMD distributions}},  in \emph{{29th International Workshop
  on Deep-Inelastic Scattering and Related Subjects}}, 8, 2022,
  \href{https://arxiv.org/abs/2208.08446}{{\ttfamily 2208.08446}}.

\bibitem{Ju:2022wia}
W.-L. Ju and M.~Sch\"onherr, \emph{{Projected transverse momentum resummation
  in top-antitop pair production at LHC}},
  \href{https://doi.org/10.1007/JHEP02(2023)075}{\emph{JHEP} {\bfseries 02}
  (2023) 075}, [\href{https://arxiv.org/abs/2210.09272}{{\ttfamily
  2210.09272}}].

\bibitem{Zhang:2020onw}
C.~Zhang, Q.-S. Dai and D.~Y. Shao, \emph{{Azimuthal decorrelation for photon
  induced dijet production in ultra-peripheral collisions of heavy ions}},
  \href{https://doi.org/10.1007/JHEP02(2023)002}{\emph{JHEP} {\bfseries 23}
  (2023) 002}, [\href{https://arxiv.org/abs/2211.07071}{{\ttfamily
  2211.07071}}].

\bibitem{Shao:2023zge}
D.~Y. Shao, C.~Zhang, J.~Zhou and Y.~J. Zhou, \emph{{Lepton pair production in
  UPCs: towards the precision test of the resummation formalism}},
  \href{https://arxiv.org/abs/2306.02337}{{\ttfamily 2306.02337}}.

\bibitem{Alrashed:2021csd}
M.~Alrashed, D.~Anderle, Z.-B. Kang, J.~Terry and H.~Xing,
  \emph{{Three-dimensional imaging in nuclei}},
  \href{https://doi.org/10.1103/PhysRevLett.129.242001}{\emph{Phys. Rev. Lett.}
  {\bfseries 129} (2022) 242001},
  [\href{https://arxiv.org/abs/2107.12401}{{\ttfamily 2107.12401}}].

\bibitem{Barry:2023qqh}
P.~C. Barry, L.~Gamberg, W.~Melnitchouk, E.~Moffat, D.~Pitonyak, A.~Prokudin
  et~al., \emph{{Tomography of pions and protons via transverse momentum
  dependent distributions}},
  \href{https://arxiv.org/abs/2302.01192}{{\ttfamily 2302.01192}}.

\bibitem{Mueller:2013wwa}
A.~H. Mueller, B.-W. Xiao and F.~Yuan, \emph{{Sudakov double logarithms
  resummation in hard processes in the small-x saturation formalism}},
  \href{https://doi.org/10.1103/PhysRevD.88.114010}{\emph{Phys. Rev. D}
  {\bfseries 88} (2013) 114010},
  [\href{https://arxiv.org/abs/1308.2993}{{\ttfamily 1308.2993}}].

\bibitem{Taels:2022tza}
P.~Taels, T.~Altinoluk, G.~Beuf and C.~Marquet, \emph{{Dijet photoproduction at
  low x at next-to-leading order and its back-to-back limit}},
  \href{https://doi.org/10.1007/JHEP10(2022)184}{\emph{JHEP} {\bfseries 10}
  (2022) 184}, [\href{https://arxiv.org/abs/2204.11650}{{\ttfamily
  2204.11650}}].

\bibitem{Caucal:2023nci}
P.~Caucal, F.~Salazar, B.~Schenke, T.~Stebel and R.~Venugopalan,
  \emph{{Back-to-back inclusive dijets in DIS at small $x$: Gluon
  Weizs\"acker-Williams distribution at NLO}},
  \href{https://arxiv.org/abs/2304.03304}{{\ttfamily 2304.03304}}.

\bibitem{CMS:2011hzb}
{\scshape CMS} collaboration, V.~Khachatryan et~al., \emph{{Dijet Azimuthal
  Decorrelations in $pp$ Collisions at $\sqrt{s} = 7$ TeV}},
  \href{https://doi.org/10.1103/PhysRevLett.106.122003}{\emph{Phys. Rev. Lett.}
  {\bfseries 106} (2011) 122003},
  [\href{https://arxiv.org/abs/1101.5029}{{\ttfamily 1101.5029}}].

\bibitem{CMS:2014qvs}
{\scshape CMS} collaboration, S.~Chatrchyan et~al., \emph{{Studies of dijet
  transverse momentum balance and pseudorapidity distributions in pPb
  collisions at $\sqrt{s_{\mathrm{NN}}} = 5.02$ $\,\text {TeV}$}},
  \href{https://doi.org/10.1140/epjc/s10052-014-2951-y}{\emph{Eur. Phys. J. C}
  {\bfseries 74} (2014) 2951},
  [\href{https://arxiv.org/abs/1401.4433}{{\ttfamily 1401.4433}}].

\bibitem{CMS:2018jpl}
{\scshape CMS} collaboration, A.~M. Sirunyan et~al., \emph{{Constraining gluon
  distributions in nuclei using dijets in proton-proton and proton-lead
  collisions at $\sqrt{s_{_\mathrm{NN}}} =$ 5.02 TeV}},
  \href{https://doi.org/10.1103/PhysRevLett.121.062002}{\emph{Phys. Rev. Lett.}
  {\bfseries 121} (2018) 062002},
  [\href{https://arxiv.org/abs/1805.04736}{{\ttfamily 1805.04736}}].

\bibitem{Eskola:2016oht}
K.~J. Eskola, P.~Paakkinen, H.~Paukkunen and C.~A. Salgado, \emph{{EPPS16:
  Nuclear parton distributions with LHC data}},
  \href{https://doi.org/10.1140/epjc/s10052-017-4725-9}{\emph{Eur. Phys. J. C}
  {\bfseries 77} (2017) 163},
  [\href{https://arxiv.org/abs/1612.05741}{{\ttfamily 1612.05741}}].

\bibitem{AbdulKhalek:2022fyi}
R.~Abdul~Khalek, R.~Gauld, T.~Giani, E.~R. Nocera, T.~R. Rabemananjara and
  J.~Rojo, \emph{{nNNPDF3.0: evidence for a modified partonic structure in
  heavy nuclei}},
  \href{https://doi.org/10.1140/epjc/s10052-022-10417-7}{\emph{Eur. Phys. J. C}
  {\bfseries 82} (2022) 507},
  [\href{https://arxiv.org/abs/2201.12363}{{\ttfamily 2201.12363}}].

\bibitem{Bauer:2000yr}
C.~W. Bauer, S.~Fleming, D.~Pirjol and I.~W. Stewart, \emph{{An Effective field
  theory for collinear and soft gluons: Heavy to light decays}},
  \href{https://doi.org/10.1103/PhysRevD.63.114020}{\emph{Phys. Rev.}
  {\bfseries D63} (2001) 114020},
  [\href{https://arxiv.org/abs/hep-ph/0011336}{{\ttfamily hep-ph/0011336}}].

\bibitem{Bauer:2001ct}
C.~W. Bauer and I.~W. Stewart, \emph{{Invariant operators in collinear
  effective theory}},
  \href{https://doi.org/10.1016/S0370-2693(01)00902-9}{\emph{Phys. Lett.}
  {\bfseries B516} (2001) 134--142},
  [\href{https://arxiv.org/abs/hep-ph/0107001}{{\ttfamily hep-ph/0107001}}].

\bibitem{Bauer:2001yt}
C.~W. Bauer, D.~Pirjol and I.~W. Stewart, \emph{{Soft collinear factorization
  in effective field theory}},
  \href{https://doi.org/10.1103/PhysRevD.65.054022}{\emph{Phys. Rev.}
  {\bfseries D65} (2002) 054022},
  [\href{https://arxiv.org/abs/hep-ph/0109045}{{\ttfamily hep-ph/0109045}}].

\bibitem{Bauer:2002nz}
C.~W. Bauer, S.~Fleming, D.~Pirjol, I.~Z. Rothstein and I.~W. Stewart,
  \emph{{Hard scattering factorization from effective field theory}},
  \href{https://doi.org/10.1103/PhysRevD.66.014017}{\emph{Phys. Rev.}
  {\bfseries D66} (2002) 014017},
  [\href{https://arxiv.org/abs/hep-ph/0202088}{{\ttfamily hep-ph/0202088}}].

\bibitem{Beneke:2002ph}
M.~Beneke, A.~P. Chapovsky, M.~Diehl and T.~Feldmann, \emph{{Soft collinear
  effective theory and heavy to light currents beyond leading power}},
  \href{https://doi.org/10.1016/S0550-3213(02)00687-9}{\emph{Nucl. Phys.}
  {\bfseries B643} (2002) 431--476},
  [\href{https://arxiv.org/abs/hep-ph/0206152}{{\ttfamily hep-ph/0206152}}].

\bibitem{Collins:1981uk}
J.~C. Collins and D.~E. Soper, \emph{{Back-To-Back Jets in QCD}},
  \href{https://doi.org/10.1016/0550-3213(81)90339-4}{\emph{Nucl. Phys.}
  {\bfseries B193} (1981) 381}.

\bibitem{Collins:1984kg}
J.~C. Collins, D.~E. Soper and G.~F. Sterman, \emph{{Transverse Momentum
  Distribution in Drell-Yan Pair and W and Z Boson Production}},
  \href{https://doi.org/10.1016/0550-3213(85)90479-1}{\emph{Nucl. Phys. B}
  {\bfseries 250} (1985) 199--224}.

\bibitem{Becher:2010tm}
T.~Becher and M.~Neubert, \emph{{{Drell-Yan} Production at Small $q_T$,
  Transverse Parton Distributions and the Collinear Anomaly}},
  \href{https://doi.org/10.1140/epjc/s10052-011-1665-7}{\emph{Eur. Phys. J.}
  {\bfseries C71} (2011) 1665},
  [\href{https://arxiv.org/abs/1007.4005}{{\ttfamily 1007.4005}}].

\bibitem{Becher:2011xn}
T.~Becher, M.~Neubert and D.~Wilhelm, \emph{{Electroweak Gauge-Boson Production
  at Small $q_T$: Infrared Safety from the Collinear Anomaly}},
  \href{https://doi.org/10.1007/JHEP02(2012)124}{\emph{JHEP} {\bfseries 02}
  (2012) 124}, [\href{https://arxiv.org/abs/1109.6027}{{\ttfamily 1109.6027}}].

\bibitem{Collins:2007nk}
J.~Collins and J.-W. Qiu, \emph{{$k_{T}$ factorization is violated in
  production of high-transverse-momentum particles in hadron-hadron
  collisions}}, \href{https://doi.org/10.1103/PhysRevD.75.114014}{\emph{Phys.
  Rev.} {\bfseries D75} (2007) 114014},
  [\href{https://arxiv.org/abs/0705.2141}{{\ttfamily 0705.2141}}].

\bibitem{Rogers:2010dm}
T.~C. Rogers and P.~J. Mulders, \emph{{No Generalized TMD-Factorization in
  Hadro-Production of High Transverse Momentum Hadrons}},
  \href{https://doi.org/10.1103/PhysRevD.81.094006}{\emph{Phys. Rev. D}
  {\bfseries 81} (2010) 094006},
  [\href{https://arxiv.org/abs/1001.2977}{{\ttfamily 1001.2977}}].

\bibitem{Catani:2011st}
S.~Catani, D.~de~Florian and G.~Rodrigo, \emph{{Space-like (versus time-like)
  collinear limits in QCD: Is factorization violated?}},
  \href{https://doi.org/10.1007/JHEP07(2012)026}{\emph{JHEP} {\bfseries 07}
  (2012) 026}, [\href{https://arxiv.org/abs/1112.4405}{{\ttfamily 1112.4405}}].

\bibitem{Forshaw:2012bi}
J.~R. Forshaw, M.~H. Seymour and A.~Siodmok, \emph{{On the Breaking of
  Collinear Factorization in QCD}},
  \href{https://doi.org/10.1007/JHEP11(2012)066}{\emph{JHEP} {\bfseries 11}
  (2012) 066}, [\href{https://arxiv.org/abs/1206.6363}{{\ttfamily 1206.6363}}].

\bibitem{Stewart:notes}
I.~W. Stewart, \emph{{Lectures on the Soft-Collinear Effective Theory}},
  {\emph{\href{http://ocw.mit.edu/courses/physics/8-851-effective-field-theory-spring-2013/lecture-notes/MIT8_851S13_scetnotes.pdf}{MIT
  Open Course Ware, Effective Field Theory (Spring 2013)}} }.

\bibitem{Becher:2014oda}
T.~Becher, A.~Broggio and A.~Ferroglia, \emph{{Introduction to Soft-Collinear
  Effective Theory}},
  \href{https://doi.org/10.1007/978-3-319-14848-9}{\emph{Lect. Notes Phys.}
  {\bfseries 896} (2015) pp.1--206},
  [\href{https://arxiv.org/abs/1410.1892}{{\ttfamily 1410.1892}}].

\bibitem{Schwartz:2013pla}
M.~D. Schwartz, \emph{{Quantum Field Theory and the Standard Model}}.
\newblock Cambridge University Press, 3, 2014.

\bibitem{Gao:2022bzi}
A.~Gao, J.~K.~L. Michel, I.~W. Stewart and Z.~Sun, \emph{{Better angle on
  hadron transverse momentum distributions at the Electron-Ion Collider}},
  \href{https://doi.org/10.1103/PhysRevD.107.L091504}{\emph{Phys. Rev. D}
  {\bfseries 107} (2023) L091504},
  [\href{https://arxiv.org/abs/2209.11211}{{\ttfamily 2209.11211}}].

\bibitem{Boussarie:2023izj}
R.~Boussarie et~al., \emph{{TMD Handbook}},
  \href{https://arxiv.org/abs/2304.03302}{{\ttfamily 2304.03302}}.

\bibitem{Collins:2011zzd}
J.~Collins, \emph{{Foundations of perturbative QCD}}, vol.~32.
\newblock Cambridge University Press, 11, 2013.

\bibitem{Catani:1996jh}
S.~Catani and M.~H. Seymour, \emph{{The Dipole formalism for the calculation of
  QCD jet cross-sections at next-to-leading order}},
  \href{https://doi.org/10.1016/0370-2693(96)00425-X}{\emph{Phys. Lett. B}
  {\bfseries 378} (1996) 287--301},
  [\href{https://arxiv.org/abs/hep-ph/9602277}{{\ttfamily hep-ph/9602277}}].

\bibitem{Kelley:2010fn}
R.~Kelley and M.~D. Schwartz, \emph{{1-loop matching and NNLL resummation for
  all partonic 2 to 2 processes in QCD}},
  \href{https://doi.org/10.1103/PhysRevD.83.045022}{\emph{Phys. Rev. D}
  {\bfseries 83} (2011) 045022},
  [\href{https://arxiv.org/abs/1008.2759}{{\ttfamily 1008.2759}}].

\bibitem{Gao:2019ojf}
A.~J. Gao, H.~T. Li, I.~Moult and H.~X. Zhu, \emph{{Precision QCD Event Shapes
  at Hadron Colliders: The Transverse Energy-Energy Correlator in the
  Back-to-Back Limit}},
  \href{https://doi.org/10.1103/PhysRevLett.123.062001}{\emph{Phys. Rev. Lett.}
  {\bfseries 123} (2019) 062001},
  [\href{https://arxiv.org/abs/1901.04497}{{\ttfamily 1901.04497}}].

\bibitem{Ahrens:2010zv}
V.~Ahrens, A.~Ferroglia, M.~Neubert, B.~D. Pecjak and L.~L. Yang,
  \emph{{Renormalization-Group Improved Predictions for Top-Quark Pair
  Production at Hadron Colliders}},
  \href{https://doi.org/10.1007/JHEP09(2010)097}{\emph{JHEP} {\bfseries 09}
  (2010) 097}, [\href{https://arxiv.org/abs/1003.5827}{{\ttfamily 1003.5827}}].

\bibitem{Ellis:2010rwa}
S.~D. Ellis, C.~K. Vermilion, J.~R. Walsh, A.~Hornig and C.~Lee, \emph{{Jet
  Shapes and Jet Algorithms in SCET}},
  \href{https://doi.org/10.1007/JHEP11(2010)101}{\emph{JHEP} {\bfseries 11}
  (2010) 101}, [\href{https://arxiv.org/abs/1001.0014}{{\ttfamily 1001.0014}}].

\bibitem{Dasgupta:2001sh}
M.~Dasgupta and G.~P. Salam, \emph{{Resummation of nonglobal QCD observables}},
  \href{https://doi.org/10.1016/S0370-2693(01)00725-0}{\emph{Phys. Lett. B}
  {\bfseries 512} (2001) 323--330},
  [\href{https://arxiv.org/abs/hep-ph/0104277}{{\ttfamily hep-ph/0104277}}].

\bibitem{Becher:2015hka}
T.~Becher, M.~Neubert, L.~Rothen and D.~Y. Shao, \emph{{Effective Field Theory
  for Jet Processes}},
  \href{https://doi.org/10.1103/PhysRevLett.116.192001}{\emph{Phys. Rev. Lett.}
  {\bfseries 116} (2016) 192001},
  [\href{https://arxiv.org/abs/1508.06645}{{\ttfamily 1508.06645}}].

\bibitem{Becher:2016mmh}
T.~Becher, M.~Neubert, L.~Rothen and D.~Y. Shao, \emph{{Factorization and
  Resummation for Jet Processes}},
  \href{https://doi.org/10.1007/JHEP11(2016)019,
  10.1007/JHEP05(2017)154}{\emph{JHEP} {\bfseries 11} (2016) 019},
  [\href{https://arxiv.org/abs/1605.02737}{{\ttfamily 1605.02737}}].

\bibitem{Dasgupta:2002bw}
M.~Dasgupta and G.~P. Salam, \emph{{Accounting for coherence in interjet $E_T$
  flow: A Case study}},
  \href{https://doi.org/10.1088/1126-6708/2002/03/017}{\emph{JHEP} {\bfseries
  03} (2002) 017}, [\href{https://arxiv.org/abs/hep-ph/0203009}{{\ttfamily
  hep-ph/0203009}}].

\bibitem{Broggio:2014hoa}
A.~Broggio, A.~Ferroglia, B.~D. Pecjak and Z.~Zhang, \emph{{NNLO hard functions
  in massless QCD}}, \href{https://doi.org/10.1007/JHEP12(2014)005}{\emph{JHEP}
  {\bfseries 12} (2014) 005},
  [\href{https://arxiv.org/abs/1409.5294}{{\ttfamily 1409.5294}}].

\bibitem{Almelid:2015jia}
O.~Almelid, C.~Duhr and E.~Gardi, \emph{{Three-loop corrections to the soft
  anomalous dimension in multileg scattering}},
  \href{https://doi.org/10.1103/PhysRevLett.117.172002}{\emph{Phys. Rev. Lett.}
  {\bfseries 117} (2016) 172002},
  [\href{https://arxiv.org/abs/1507.00047}{{\ttfamily 1507.00047}}].

\bibitem{Almelid:2017qju}
O.~Almelid, C.~Duhr, E.~Gardi, A.~McLeod and C.~D. White, \emph{{Bootstrapping
  the QCD soft anomalous dimension}},
  \href{https://doi.org/10.1007/JHEP09(2017)073}{\emph{JHEP} {\bfseries 09}
  (2017) 073}, [\href{https://arxiv.org/abs/1706.10162}{{\ttfamily
  1706.10162}}].

\bibitem{Chiu:2011qc}
J.-y. Chiu, A.~Jain, D.~Neill and I.~Z. Rothstein, \emph{{The Rapidity
  Renormalization Group}},
  \href{https://doi.org/10.1103/PhysRevLett.108.151601}{\emph{Phys. Rev. Lett.}
  {\bfseries 108} (2012) 151601},
  [\href{https://arxiv.org/abs/1104.0881}{{\ttfamily 1104.0881}}].

\bibitem{Chiu:2012ir}
J.-Y. Chiu, A.~Jain, D.~Neill and I.~Z. Rothstein, \emph{{A Formalism for the
  Systematic Treatment of Rapidity Logarithms in Quantum Field Theory}},
  \href{https://doi.org/10.1007/JHEP05(2012)084}{\emph{JHEP} {\bfseries 05}
  (2012) 084}, [\href{https://arxiv.org/abs/1202.0814}{{\ttfamily 1202.0814}}].

\bibitem{Kang:2013wca}
Z.-B. Kang, X.~Liu, S.~Mantry and J.-W. Qiu, \emph{{Probing nuclear dynamics in
  jet production with a global event shape}},
  \href{https://doi.org/10.1103/PhysRevD.88.074020}{\emph{Phys. Rev. D}
  {\bfseries 88} (2013) 074020},
  [\href{https://arxiv.org/abs/1303.3063}{{\ttfamily 1303.3063}}].

\bibitem{Gelis:2010nm}
F.~Gelis, E.~Iancu, J.~Jalilian-Marian and R.~Venugopalan, \emph{{The Color
  Glass Condensate}},
  \href{https://doi.org/10.1146/annurev.nucl.010909.083629}{\emph{Ann. Rev.
  Nucl. Part. Sci.} {\bfseries 60} (2010) 463--489},
  [\href{https://arxiv.org/abs/1002.0333}{{\ttfamily 1002.0333}}].

\bibitem{Collins:2014jpa}
J.~Collins and T.~Rogers, \emph{{Understanding the large-distance behavior of
  transverse-momentum-dependent parton densities and the Collins-Soper
  evolution kernel}},
  \href{https://doi.org/10.1103/PhysRevD.91.074020}{\emph{Phys.Rev.} {\bfseries
  D91} (2015) 074020}, [\href{https://arxiv.org/abs/1412.3820}{{\ttfamily
  1412.3820}}].

\bibitem{Aidala:2014hva}
C.~Aidala, B.~Field, L.~Gamberg and T.~Rogers, \emph{{Limits on TMD Evolution
  From Semi-Inclusive Deep Inelastic Scattering at Moderate $Q$}},
  \href{https://doi.org/10.1103/PhysRevD.89.094002}{\emph{Phys.Rev.} {\bfseries
  D89} (2014) 094002}, [\href{https://arxiv.org/abs/1401.2654}{{\ttfamily
  1401.2654}}].

\bibitem{Sun:2014dqm}
P.~Sun, J.~Isaacson, C.~P. Yuan and F.~Yuan, \emph{{Nonperturbative functions
  for SIDIS and Drell\textendash{}Yan processes}},
  \href{https://doi.org/10.1142/S0217751X18410063}{\emph{Int. J. Mod. Phys. A}
  {\bfseries 33} (2018) 1841006},
  [\href{https://arxiv.org/abs/1406.3073}{{\ttfamily 1406.3073}}].

\bibitem{Landry:2002ix}
F.~Landry, R.~Brock, P.~M. Nadolsky and C.~P. Yuan, \emph{{Tevatron Run-1 Z
  boson data and Collins-Soper-Sterman resummation formalism}},
  \href{https://doi.org/10.1103/PhysRevD.67.073016}{\emph{Phys. Rev.}
  {\bfseries D67} (2003) 073016},
  [\href{https://arxiv.org/abs/hep-ph/0212159}{{\ttfamily hep-ph/0212159}}].

\bibitem{Konychev:2005iy}
A.~V. Konychev and P.~M. Nadolsky, \emph{{Universality of the
  Collins-Soper-Sterman nonperturbative function in gauge boson production}},
  \href{https://doi.org/10.1016/j.physletb.2005.12.063}{\emph{Phys.Lett.}
  {\bfseries B633} (2006) 710--714},
  [\href{https://arxiv.org/abs/hep-ph/0506225}{{\ttfamily hep-ph/0506225}}].

\bibitem{Bacchetta:2017gcc}
A.~Bacchetta, F.~Delcarro, C.~Pisano, M.~Radici and A.~Signori,
  \emph{{Extraction of partonic transverse momentum distributions from
  semi-inclusive deep-inelastic scattering, Drell-Yan and Z-boson production}},
  \href{https://doi.org/10.1007/JHEP06(2017)081}{\emph{JHEP} {\bfseries 06}
  (2017) 081}, [\href{https://arxiv.org/abs/1703.10157}{{\ttfamily
  1703.10157}}].

\bibitem{Bacchetta:2022awv}
{\scshape MAP} collaboration, A.~Bacchetta, V.~Bertone, C.~Bissolotti,
  G.~Bozzi, M.~Cerutti, F.~Piacenza et~al., \emph{{Unpolarized transverse
  momentum distributions from a global fit of Drell-Yan and semi-inclusive
  deep-inelastic scattering data}},
  \href{https://doi.org/10.1007/JHEP10(2022)127}{\emph{JHEP} {\bfseries 10}
  (2022) 127}, [\href{https://arxiv.org/abs/2206.07598}{{\ttfamily
  2206.07598}}].

\bibitem{Echevarria:2020hpy}
M.~G. Echevarria, Z.-B. Kang and J.~Terry, \emph{{Global analysis of the Sivers
  functions at NLO+NNLL in QCD}},
  \href{https://doi.org/10.1007/JHEP01(2021)126}{\emph{JHEP} {\bfseries 01}
  (2021) 126}, [\href{https://arxiv.org/abs/2009.10710}{{\ttfamily
  2009.10710}}].

\bibitem{Dulat:2015mca}
S.~Dulat, T.-J. Hou, J.~Gao, M.~Guzzi, J.~Huston, P.~Nadolsky et~al.,
  \emph{{New parton distribution functions from a global analysis of quantum
  chromodynamics}},
  \href{https://doi.org/10.1103/PhysRevD.93.033006}{\emph{Phys. Rev. D}
  {\bfseries 93} (2016) 033006},
  [\href{https://arxiv.org/abs/1506.07443}{{\ttfamily 1506.07443}}].

\bibitem{Mueller:2016xoc}
A.~H. Mueller, B.~Wu, B.-W. Xiao and F.~Yuan, \emph{{Medium Induced Transverse
  Momentum Broadening in Hard Processes}},
  \href{https://doi.org/10.1103/PhysRevD.95.034007}{\emph{Phys. Rev. D}
  {\bfseries 95} (2017) 034007},
  [\href{https://arxiv.org/abs/1608.07339}{{\ttfamily 1608.07339}}].

\bibitem{Cacciari:2008gp}
M.~Cacciari, G.~P. Salam and G.~Soyez, \emph{{The anti-$k_t$ jet clustering
  algorithm}}, \href{https://doi.org/10.1088/1126-6708/2008/04/063}{\emph{JHEP}
  {\bfseries 04} (2008) 063},
  [\href{https://arxiv.org/abs/0802.1189}{{\ttfamily 0802.1189}}].

\bibitem{Idilbi:2008vm}
A.~Idilbi and A.~Majumder, \emph{{Extending Soft-Collinear-Effective-Theory to
  describe hard jets in dense QCD media}},
  \href{https://doi.org/10.1103/PhysRevD.80.054022}{\emph{Phys. Rev. D}
  {\bfseries 80} (2009) 054022},
  [\href{https://arxiv.org/abs/0808.1087}{{\ttfamily 0808.1087}}].

\bibitem{Ovanesyan:2011xy}
G.~Ovanesyan and I.~Vitev, \emph{{An effective theory for jet propagation in
  dense QCD matter: jet broadening and medium-induced bremsstrahlung}},
  \href{https://doi.org/10.1007/JHEP06(2011)080}{\emph{JHEP} {\bfseries 06}
  (2011) 080}, [\href{https://arxiv.org/abs/1103.1074}{{\ttfamily 1103.1074}}].

\bibitem{Rothstein:2016bsq}
I.~Z. Rothstein and I.~W. Stewart, \emph{{An Effective Field Theory for Forward
  Scattering and Factorization Violation}},
  \href{https://doi.org/10.1007/JHEP08(2016)025}{\emph{JHEP} {\bfseries 08}
  (2016) 025}, [\href{https://arxiv.org/abs/1601.04695}{{\ttfamily
  1601.04695}}].

\bibitem{Kang:2017frl}
Z.-B. Kang, F.~Ringer and I.~Vitev, \emph{{Inclusive production of small radius
  jets in heavy-ion collisions}},
  \href{https://doi.org/10.1016/j.physletb.2017.03.067}{\emph{Phys. Lett. B}
  {\bfseries 769} (2017) 242--248},
  [\href{https://arxiv.org/abs/1701.05839}{{\ttfamily 1701.05839}}].

\bibitem{Al-Mashad:2022zbq}
M.~A. Al-Mashad, A.~van Hameren, H.~Kakkad, P.~Kotko, K.~Kutak, P.~van Mechelen
  et~al., \emph{{Dijet azimuthal correlations in p-p and p-Pb collisions at
  forward LHC calorimeters}},
  \href{https://doi.org/10.1007/JHEP12(2022)131}{\emph{JHEP} {\bfseries 12}
  (2022) 131}, [\href{https://arxiv.org/abs/2210.06613}{{\ttfamily
  2210.06613}}].

\bibitem{STAR:2023xvk}
{\scshape STAR} collaboration, \emph{{Measurement of transverse single-spin
  asymmetries for dijet production in polarized proton-proton collisions at
  $\sqrt{s}$ = 200 $\mathrm{GeV}$}},
  \href{https://arxiv.org/abs/2305.10359}{{\ttfamily 2305.10359}}.

\end{thebibliography}\endgroup

\end{document}